\documentclass[a4paper,11pt]{article}
\usepackage{jcappub} 
\usepackage{lineno}

\usepackage{blindtext}
\usepackage{cancel}
\usepackage{mathtools, mathalfa, amssymb, amsmath, amsfonts, amsthm, bm}
\usepackage{graphicx}
\usepackage{multirow}
\usepackage[table, dvipsnames]{xcolor}
\usepackage{color}
\usepackage{framed}
\usepackage{aas_macros}

\usepackage{amsthm} 

\newtheorem{example}{Example}

\usepackage{appendix}
\usepackage{float}
\usepackage{subcaption}
\usepackage{xcolor}
\usepackage{orcidlink}
\usepackage{algorithm}
\usepackage{algpseudocode}

\usepackage{hyperref}
\hypersetup{
    colorlinks = true,
    citecolor  = blue,
    linkcolor = blue,
    urlcolor  = blue,
    linkbordercolor = white
}

\graphicspath{{./figures/}}
\usepackage[bottom]{footmisc}

\usepackage{cleveref}

\DeclareMathOperator{\Ai}{Ai}

\colorlet{green}{teal}

\title{Resurgence and Hyperasymptotics in Wave Optics Astronomy}

  \author{Job Feldbrugge
  \orcidlink{0000-0003-2414-8707}}
  \affiliation{Higgs Centre for Theoretical Physics, University of Edinburgh, James Clerk Maxwell Building, Edinburgh EH9 3FD, UK}

  \author{Samuel Crew
  \orcidlink{0000-0002-6136-0910}} 
  \affiliation{Department of Physics, National Tsing Hua University, Hsinchu, Taiwan.}

  \author{Ue-Li Pen \orcidlink{0000-0003-2155-9578}}
  \affiliation{Institute of Astronomy and Astrophysics, Academia Sinica, Astronomy-Mathematics Building, No. 1, Section 4, Roosevelt Road, Taipei 106319, Taiwan}
  \affiliation{Canadian Institute for Theoretical Astrophysics, University of Toronto, 60 St. George Street, Toronto, ON M5S 3H8, Canada}
  \affiliation{Perimeter Institute for Theoretical Physics, 31 Caroline St. North, Waterloo, ON, Canada N2L 2Y5}
  \affiliation{Department of Physics, University of Toronto, 60 St. George Street, Toronto, ON M5S 1A7, Canada}
  \affiliation{Dunlap Institute for Astronomy \& Astrophysics, University of Toronto, 50 St. George Street, Toronto, ON M5S 3H4, Canada}
  \affiliation{Canadian Institute for Advanced Research, CIFAR program in Gravitation and Cosmology}

\emailAdd{job.feldbrugge@ed.ac.uk}

\abstract{       
  With the discovery of gravitational waves and fast radio bursts, wave optics has become increasingly relevant in astrophysics.  This paper studies the behaviour of random gravitational and plasma lenses, presenting the refractive and diffractive expansions, with higher-order terms that allow error estimates and embody the counterintuitive {\it resurgence} phenomenon. 
        
  Specifically, we show that the diffractive expansion converges for a broad class of bounded lens models and provides an efficient description of interference patterns across frequency regimes. Next, building on Picard–Lefschetz techniques, we derive the full refractive expansion to arbitrary order, organising it into a transseries. Near caustics, the standard transseries is supplemented with uniform asymptotics. We study this transseries, with both Borel and hyperasymptotic resummation yielding systematic approximations to lensing integrals at all frequencies. Our results give a framework for modelling wave optics lensing near caustics and beyond the geometric optics approximation and thereby illustrate how tools from resurgence and asymptotic analysis can be applied to practical problems in astrophysics.

  Near caustic singularities, the post-refractive corrections diverge, while the uniform asymptotic expansion becomes accurate. We use the leading uniform approximation to derive the strong wave optics suppression of off-axis caustics, which clarifies their subdominant role.}

\begin{document}
\maketitle
\flushbottom

    \section{Introduction}
    The theory and observation of gravitational and plasma lensing play a central role in modern astronomy. Historically, most studies have relied on the geometric optics approximation, in which light is treated as rays following null geodesics in curved spacetime or as being deflected upon traversing a lens plane. While geometric optics remains adequate for most purposes, wave optics has become increasingly important with the advent of observations of gravitational waves, fast radio bursts, and pulsars \cite{LIGOScientific:2016, Main:2018, Petroff:2019}. In these contexts—where coherent, long-wavelength radiation is lensed—the wave nature of light can be directly probed through observable interference effects. This is especially relevant near caustics, where the intensity is amplified and the geometric optics approximation breaks down. The study of wave optics thus lies at the fertile intersection of caustics, singularity theory, and quantum interference. Although our present work centers on lens integrals, the results hold more general significance for the analysis of multidimensional oscillatory integrals. In particular, we anticipate future applications to real-time Feynman path integrals.

    Wave effects in lensing constitute a rich phenomenon that is notoriously challenging to model: the governing Fresnel–Kirchhoff integrals are highly oscillatory, and their convergence is delicate. This complexity has historically restricted detailed calculations to the simplest case: lensing by a single, isolated point mass \cite{Nakamura:1999}. Recently, two of the authors developed an efficient numerical method based on the stationary-phase approach and Picard–Lefschetz theory \cite{Feldbrugge:2023b}. By deforming the original integration domain onto a set of complex Lefschetz thimbles—each associated with a saddle point and justified by Cauchy’s theorem—the conditionally convergent integral is transformed into a sum of absolutely convergent contributions. This method has already enabled the systematic study of several wave-optics lens models \cite{Feldbrugge:2020, Jow:2020rcy, Jow:2021, Jow:2021gnv, Bonga:2025, Jow:2023, Feldbrugge:2023c}. In this paper, we, for the first time, extend the analysis of wave optics in astrophysics to the analytical realm using \textit{resurgence theory}.

    In this paper, we prove that the \textit{diffractive expansion} -- normally developed in the diffractive regime -- converges for any frequency and allows us to explore the full interference pattern and the caustics of the lens system in both the diffractive and refractive regimes without any knowledge of classical rays. Next, we provide a new derivation of the \textit{refractive expansion} for $d$-dimensional lens integrals to arbitrary order, leading to an algorithm for its evaluation. So far, studies of the refractive expansion have been limited to the eikonal approximation. However, the full refractive expansion of the Kirchhoff-Fresnel integral turns out to diverge, forming a transseries consisting of several asymptotic series. Using the \textit{uniform approximation, the superasymptotic and the hyperasymptotic approximation} techniques from resurgence theory, we show how to approximate the lens integral with different levels of precision using the formal refractive expansion. In particular, the hyperasymptotic approximation of the refractive expansion allows us to study the lens integral in both the refractive and diffractive regimes with arbitrary precision. Consequently, the diffractive and refractive expansions together form a powerful analytic framework to model the lens integral for the full frequency range.

    While resurgence is a well-developed and sophisticated field of mathematics, this is the first application to lensing in wave optics. At present, most studies in resurgence either target specific mathematical aspects of the theory or are limited to one-dimensional integrals. In this paper, we provide a pedagogical account of the use of resurgence theory to study physical problems and, in the process, derive several new results in resurgence relevant to the evaluation of multidimensional oscillatory integrals. The central mathematical points are demonstrated with a series of elementary examples.

     \subsection*{Outline.}
     The paper is structured as follows. We begin in section \ref{sec:diffractiveregime} with a discussion of the diffractive expansion. In section \ref{sec:refractiveregime}, we turn to the refractive expansion and derive the associated formal transseries that encodes the formal asymptotic approximation. In \cref{sec:resurgence} use resurgence theory to extract numerical values for the lens integral using the transseries. We conclude in \cref{sec:conclusion} with a discussion and outlook for future investigations.

    \section{The diffractive expansion}\label{sec:diffractiveregime}
\begin{figure}
        \centering
        \includegraphics[width=0.7\textwidth]{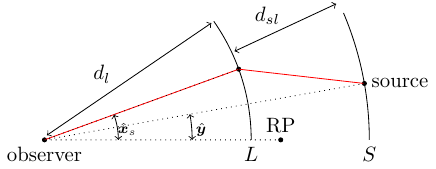}
        \caption{The lens system displaying a ray moving from the angular position $\hat{\bm{y}}$ on the source plane $S$ to the angular position $\hat{\bm{x}}$ on the lens plane $L$ to the observer.}\label{fig:lensSystem}
    \end{figure}

    In the present study, we develop the diffractive expansion; the refractive expansion; and resurgence theory for the lensing of radiation by a single thin lens. Let us consider a thin lens system, modelled by the dimensionless Kirchhoff-Fresnel diffraction integral \cite{Fresnel:1816, Kirchhoff:1882}, 
    \begin{align}
        \Psi(\bm{y}) = \left(\frac{\omega}{2 \pi i}\right)^{d/2} \int e^{i \omega T(\bm{x},\bm{y})}\mathrm{d}\bm{x}\,,\label{eq:KF}
    \end{align}
    in terms of the dimensionless frequency $\omega$; the time delay 
    \begin{align}
        T(\bm{x},\bm{y}) = \frac{(\bm{x}-\bm{y})^2}{2} + \varphi(\bm{x})\,;
    \end{align}
    the phase variation induced by the lens $\varphi$ and the dimension of the lens $d$. The modulus squared of the lens amplitude $\Psi(\bm{y})$ is interpreted as the intensity $I(\bm{y}) = |\Psi(\bm{y})|^2$ at a point $\bm{y}$ in the image plane (see \cref{fig:lensSystem} for a diagram of the thin lens system with dimensional parameters.). The Kirchhoff-Fresnel integral ranges over all possible rays moving from the source to the observer via the lens plane at point $\bm{x}$. For convenience, we use dimensionless units with the dimensionless angular position on the lens plane $\bm{x}=\hat{\bm{x}}/\theta_*, \bm{y} = \hat{\bm{y}}/\theta_*$ with some angular scale $\theta_*$ associated with the lens system, and the dimensionless frequency $\omega = \frac{\theta_*^2 d_s}{c d_{sl} d_l} \nu$  with the angular frequency of the radiation $\nu$. For a detailed derivation of the Kirchhoff-Fresnel integral from the real-time path integral, see \cite{Schneider:1992, Nakamura:1999,Feldbrugge:2023b} and we refer to \cite{Braga:2024} for a related derivation of gravitational lensing using relativistic world-line quantisation.

    In the diffractive expansion, typically used in the diffractive regime $\omega \ll 1$, we expand the exponential of the phase variation
    \begin{align}
        e^{i \omega \varphi(\bm{x})} = \sum_{n=0}^\infty \frac{1}{n!}[i \omega \varphi(\bm{x})]^n\,,
    \end{align}
    and commute the infinite sum with the integration symbol to obtain the expansion
    \begin{align}\label{eq:diffractiveexpansion}
        \Psi(\bm{y}) &= \left(\frac{\omega}{2 \pi i}\right)^{d/2} \int
        e^{i \omega \frac{(\bm{x}-\bm{y})^2}{2} }
        \sum_{n=0}^\infty \frac{1}{n!} \left[i \omega \varphi(\bm{x})\right]^n
        \mathrm{d}\bm{x}\\
        &\sim  \left(\frac{\omega}{2 \pi i}\right)^{d/2} \sum_{n=0}^\infty \frac{\omega^n}{n!} \int
        e^{i \omega \frac{(\bm{x}-\bm{y})^2}{2} }
         \left[i  \varphi(\bm{x})\right]^n
        \mathrm{d}\bm{x}\\
        &= \sum_{n=0}^\infty a_n(\bm{y}) \omega^n\,,\label{eq:diffractive}
    \end{align}
    with coefficients
    \begin{align}
        a_n(\bm{y}) = \left(\frac{\omega}{2 \pi i}\right)^{d/2} \frac{1}{n!} \int_{\mathbb{R}^d} e^{i \omega\frac{(\bm{x}- \bm{y})^2}{2}} \left[i \varphi(\bm{x})\right]^n\mathrm{d}\bm{x}\,.\label{eq:coeff_a}
    \end{align}
    By construction, the first coefficient $a_0=1$ while the subsequent coefficients $a_n$ are functions of the frequency. The diffractive expansion is typically used in the diffractive regime -- for small frequencies $\omega$ -- where the expansion typically converges rapidly to the lens amplitude.
    
    Remarkably, the diffractive expansion \eqref{eq:diffractive} does not necessarily converge as the lens integral converges only conditionally!\footnote{
        The integral $\int_\Omega f(x) \mathrm{d}x$ of the function $f$ over the domain $\Omega$ converges absolutely when the integral over the modulus converges \textit{i.e.}, $\int_\Omega | f(x)|\mathrm{d}x <  \infty$. When the integral converges, but the integral over its modulus diverges, the integral is conditionally convergent. The Kirchhoff-Fresnel integral is an example of a conditionally convergent as the lens amplitude $\Psi(\bm{y})$ is finite but the integral over the modulus diverges, \textit{i.e.}, $\int \left|e^{i \omega T(\bm{x},\bm{y})}\right|\mathrm{d}\bm{x}= \int \mathrm{d}\bm{x} = \infty$.} 
        When moving the infinite sum outside the integration symbol, we are not, by the dominated convergence theorem,\footnote{
        \textit{The dominated convergence theorem:} The limit of a set of integrals is guaranteed to converge to the integral of the limit when there exists a measurable function dominating the integrands of the sequence. Explicitly, $\lim_{n\to \infty} \int f_n(x)\mathrm{d}x =  \int \lim_{n\to \infty} f_n(x)\mathrm{d}x$ for sequence $\{f_n\}_{n=1}^\infty$ when there exists a function $g$ such that $\int|g(\bm{x})|\mathrm{d}\bm{x} < \infty$ and $f_n(\bm{x}) < g(\bm{x})$ for all $\bm{x}$ and $n$.
        } guaranteed to recover the lens integral.
    To illustrate this phenomenon, let us study the diffractive expansion of a closely related integral.
    \begin{example}\label{ex:quartic}
        Consider the one-dimensional quartic oscillatory integral
        \begin{align}
            \Phi &= \sqrt{\frac{\omega}{\pi i}} \int_{-\infty}^\infty e^{i\omega (x^2 + x^4)}\mathrm{d}x \\
            &= \sqrt{\frac{\omega}{4\pi i}} e^{-\frac{i \omega}{8}} K_{1/4}\left(-\frac{i \omega}{8}\right)\,,
        \end{align}
        expressed in terms of the modified Bessel function of the second kind $K_n(x)$. Expanding $e^{i \omega x^4} = \sum_{n=0}^\infty \frac{(i \omega)^2}{n!} x^{4n}$ and moving the sum outside the integral yields the power series
        \begin{align}
            \Phi 
            \sim \sqrt{\frac{\omega}{\pi i}} \sum_{n=0}^\infty  \frac{(i \omega)^n}{n!} \int_{-\infty}^\infty e^{i\omega x^2}  x^{4n}\mathrm{d}x
            =\sum_{n=0}^\infty \frac{(4 n - 1)!!}{4^{ n} n! i^n \omega^n} \,,
            \label{eq:quartic_expansion}
        \end{align}
        with the double factorial $n!!=n (n-2)\cdots$. The partial sum
        \begin{align}
            \Phi_N = \sum_{n=0}^N \frac{(4 n - 1)!!}{4^{ n} n! i^n \omega^n},
        \end{align}
        first approaches the value of the quartic integral before eventually diverging (see \cref{fig:asymptotic}). The terms in the expansion grow factorially with $n$. Note that after evaluating the coefficients $a_n$, the expansion organizes itself in inverse powers of the frequency.

        \begin{figure}
            \centering
            \begin{subfigure}[b]{0.45\textwidth}
                \includegraphics[width=\textwidth]{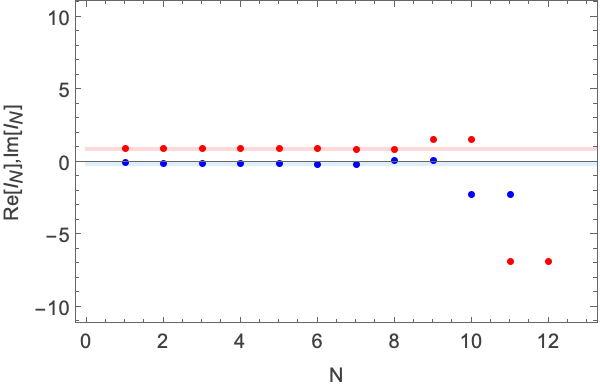}
                \caption{The real (red) and imaginary parts (blue) of the exact integral $I$ (the lines) compared with the partial sum $I_N$ (the points).}
            \end{subfigure}
            \begin{subfigure}[b]{0.45\textwidth}
                \includegraphics[width=\textwidth]{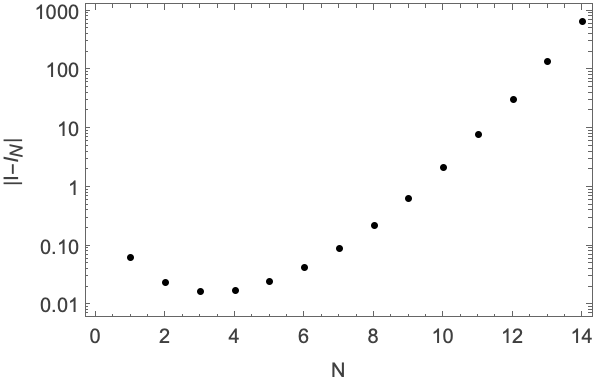}
                \caption{The error of the partial sum $I_N$ with respect to the exact result $I$.\phantom{test} \phantom{test} \phantom{test} \phantom{test} \phantom{test} \phantom{test} \phantom{test} \phantom{test} \phantom{test} \phantom{test} \phantom{test} }
            \end{subfigure}
            \caption{The partial sum $I_N$ of the asymptotic series of the integral compared with the closed form expression for $I$ for $\omega = 10$ as a function of the truncation point $N$.}\label{fig:asymptotic}
        \end{figure}
    \end{example}

    Such a divergent series approximation is referred to as an asymptotic series and occurs in many physical problems. Consider, for example, the famous Dyson series in quantum electrodynamics or Feynman diagram expansions in quantum field theory more generally. Dyson argued that if the perturbative series in the electric charge $e$ would converge for small $e$, it would define an analytic function in a small open disk $|e^2| <R$ in the complex plane, making the power series also converge for $e^2<0$. However, this is unphysical as negative $e^2$ would render the vacuum unstable \cite{Dyson:1952}. From the path integral perspective, the Feynman diagram expansion mirrors the diffractive expansion, and its asymptotic nature arises from moving an infinite sum outside of a conditionally convergent functional integral. Throughout history, asymptotic series intrigued and challenged some of the greatest mathematicians \cite{berry2017divergent}. As we will see in \cref{sec:resurgence}, in more recent decades, the mathematical field of resurgence has developed a sophisticated understanding of asymptotic series and moreover has developed concrete methods to extract from them increasingly accurate approximations of integrals and solution to differential equations.

    However, to our surprise, the diffractive expansion for the vast majority of lens systems -- despite relying on the exchange of the sum and integration symbol for a conditionally convergent integral -- does converge to the Kirchhoff-Fresnel integral for any frequency. Specifically, when the integral over the phase-variation $\varphi$ is integrable and bounded, the diffractive expansion converges. To see this, write $\varphi$ as $\alpha \phi$ with $\alpha >0$ such that $|\phi(\bm{x})| <1$ for all $\bm{x}$. The coefficients in the series expansion are bounded by a rapidly decreasing function for large $n$,
    \begin{align}
        |a_n \omega^n| &=  \left|\left(\frac{\omega}{2 \pi i}\right)^{d/2}  \frac{\omega^n}{n!} \int
        e^{i \omega \frac{(\bm{x}-\bm{y})^2}{2} }
            \left[i  \varphi(\bm{x})\right]^n
        \mathrm{d}\bm{x} \right|\\
        &\leq
        \left(\frac{\omega}{2 \pi }\right)^{d/2}  \frac{(\alpha \omega)^n}{n!} \int
            \left|  \phi(\bm{x})\right|^n
        \mathrm{d}\bm{x} \\
        &\leq
        \left(\frac{\omega}{2 \pi }\right)^{d/2} \frac{(\alpha \omega)^n}{n!} \Omega\,,
    \end{align}
    as $\int |\phi(\bm{x})|^n \mathrm{d}\bm{x} \leq  \int |\phi(\bm{x})|\mathrm{d}\bm{x} = \Omega$ since $|\phi(\bm{x})| < 1$. Cauchy's ratio test\footnote{
        \textit{Cauchy's ratio test:} The series $\sum_{n=1}^\infty a_n$ converges absolutely when the limit $L= \lim_{n \to \infty} \left|\frac{a_{n+1}}{a_n}\right|$ converges to a number $L <1$. The series diverges when $L>1$. When $L=1$ or the limit fails to exist, the Cauchy's ratio test is inconclusive.}
    then guarantees the convergence of the diffractive expansion as the ratio of subsequent terms in the series vanishes in the limit
    \begin{align}
        \lim_{n \to \infty} \left|\frac{a_{n+1} \omega^{n+1}}{a_n \omega^n}\right| 
        = 
        \lim_{n \to \infty}\frac{\alpha \omega^2 }{n+1} = 0\,.
    \end{align}
    To the best of our knowledge, the convergence of the diffractive expansion, irrespective of the frequency, for bounded and integrable lenses is a new result. The diffractive expansion allows us to explore both the diffractive and the refractive regime in wave optics and analyse the nature of caustics and interference patterns without any knowledge of either real or complex classical rays! The natural speculation is that the asymptotic nature of Feynman diagram expansions results from the polynomial nature of the interaction terms.

\begin{figure}
        \centering
        \includegraphics[width=0.5\textwidth]{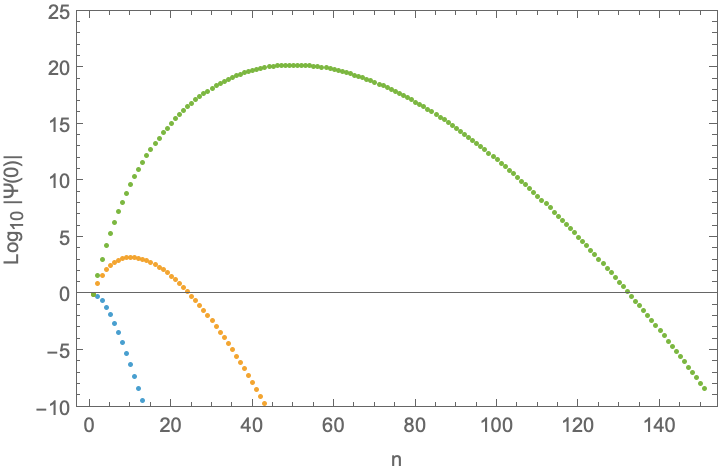}
        \caption{The convergence of the diffractive expansion for the Lorentzian lens model at $x=0$ for $\omega=1,10,50$ (blue, yellow, green).}\label{fig:refractive_convergence}
    \end{figure}

    Interestingly, the diffractive expansion evolves in precisely the opposite manner to the asymptotic expansion. The terms $a_n \omega^n$, scaling like $\frac{(\alpha\, \omega)^n}{n!}$, first rapidly grow peaking around $n=\lfloor \alpha \, \omega \rfloor$ -- the integer part of $\alpha\, \omega$ -- before decaying to zero (see \cref{fig:refractive_convergence}). As we will see, asymptotic series typically scale like $n!/F^n$ for some constant $F$. For small frequencies, the diffractive expansion is often highly efficient but in the large frequency regime, we require many terms to closely approximate the lens integral. Moreover, the interference pattern emerges from the sum of increasingly large terms, requiring extremely precise evaluations of the coefficients $a_n(\bm{y})$, making the expansion numerically unstable at high frequencies thereby motivating the development of alternative methods for the high frequency regime (see \cref{sec:refractiveregime} and \cref{sec:resurgence}). Nonetheless, the diffractive expansion provides a direct method for the evaluation of lens integrals at any frequency using the Gaussian lens.
    
    \bigskip
    For general lens models, evaluating the coefficients $a_n$ to the required accuracy is computationally the most expensive step for evaluating the diffractive expansion at high frequency. To circumvent this problem, let us consider the diffractive expansion for the Gaussian lens:

    \begin{example}
        For the Gaussian lens, the lens amplitude is 
        \begin{align}
            \Psi(\bm{y}) 
            &= \left(\frac{\omega}{2 \pi i}\right)^{d/2} \int e^{i \omega \left[\frac{(\bm{x}-\bm{y})^2}{2} + \alpha \exp\left[-\frac{1}{2}(\bm{x}-\bm{\mu})^T \Sigma^{-1} (\bm{x}-\bm{\mu})\right]\right]}\mathrm{d}\bm{x}\\
            &= \left(\frac{\omega}{2 \pi i}\right)^{d/2} \sum_{n=0}^\infty \frac{(i \omega \alpha)^n}{n!}\int e^{i \omega \frac{(\bm{x}-\bm{y})^2}{2}  -\frac{n}{2}(\bm{x}-\bm{\mu})^T \Sigma^{-1} (\bm{x}-\bm{\mu})}\mathrm{d}\bm{x}\\
            &=\sum_{n=0}^\infty a_n (\bm{y})\omega^n\,,
        \end{align} 
        in terms of the amplitude $\alpha$, the mean $\bm{\mu}$ and the covariance $\Sigma$. The coefficients of the diffractive expansion assume the form
        \begin{align}
            a_n(\bm{y}) =
            \frac{(i \alpha)^n}{n!}\frac{\left(-i \omega\right)^{d/2}}{\sqrt{\det (n \Sigma^{-1} - i \omega I)}}
            e^{-\frac{\omega^2}{2} (\bm{y} - \bm{\mu})^T(n \Sigma^{-1} - i \omega I)^{-1} (\bm{y} - \bm{\mu}) + i \omega \frac{(\bm{y} - \bm{\mu})^2}{2}}\,.
        \end{align}
    \end{example}
    
    \begin{figure}
        \centering
        \begin{subfigure}[b]{0.32\textwidth}
            \includegraphics[width=\textwidth]{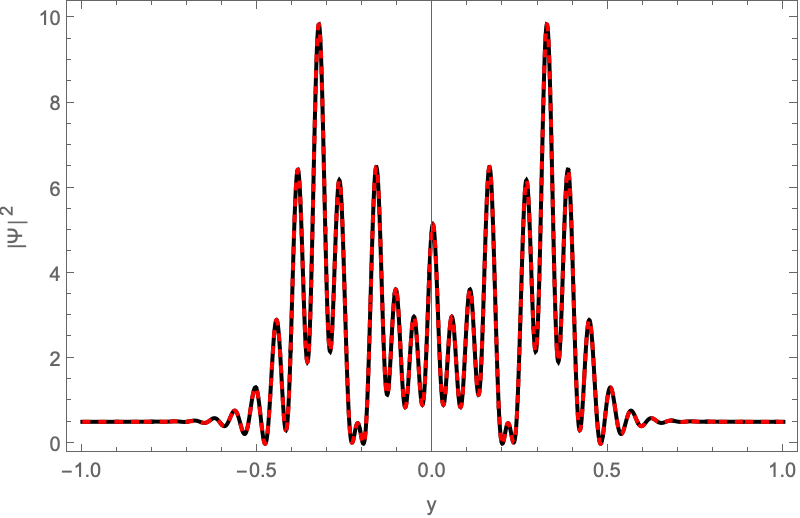}
            \caption{$\omega=50$}
        \end{subfigure}
        \begin{subfigure}[b]{0.32\textwidth}
            \includegraphics[width=\textwidth]{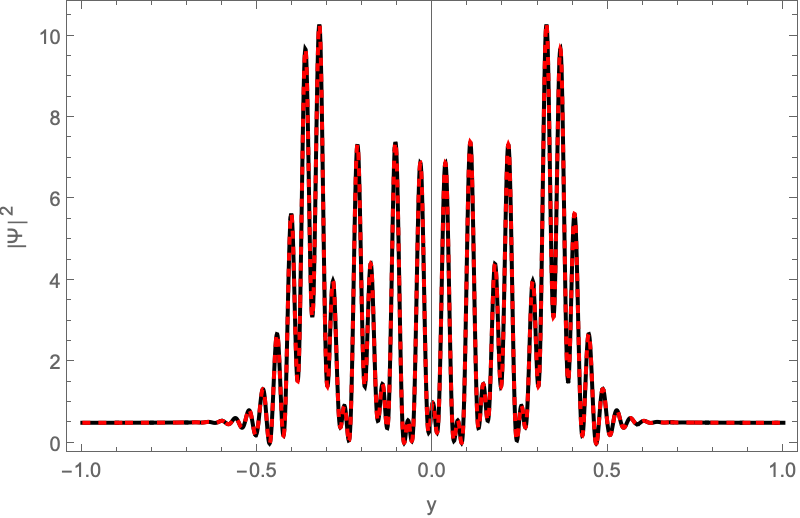}
            \caption{$\omega=75$}
        \end{subfigure}
        \begin{subfigure}[b]{0.32\textwidth}
            \includegraphics[width=\textwidth]{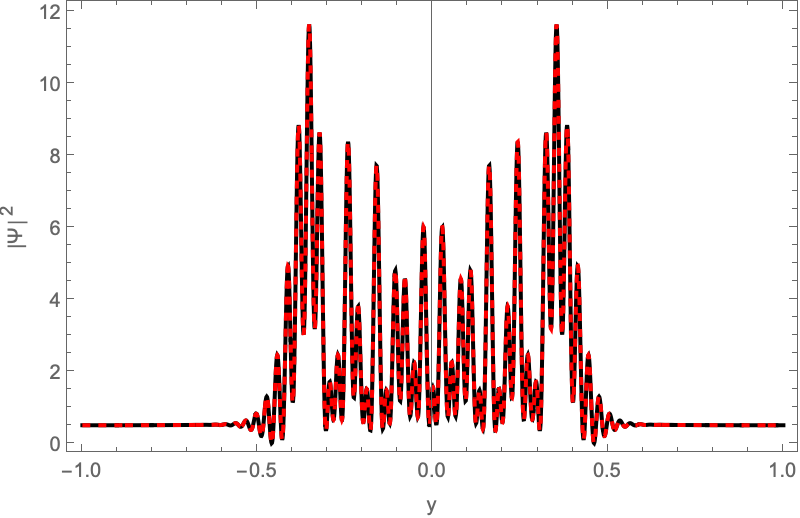}
            \caption{$\omega=100$}
        \end{subfigure}
        \caption{The interference pattern $|I(y)=|\Psi(y)|^2$ for the one-dimensional Gaussian lens model with the parameters $\alpha =2, \mu = 0$, and $\sigma = 1$ evaluated with Picard-Lefschetz theory (black) and the diffraction expansion (red) truncated at $N=550$.}\label{fig:Gaussian}

        \begin{subfigure}[b]{0.32\textwidth}
            \includegraphics[width=\textwidth]{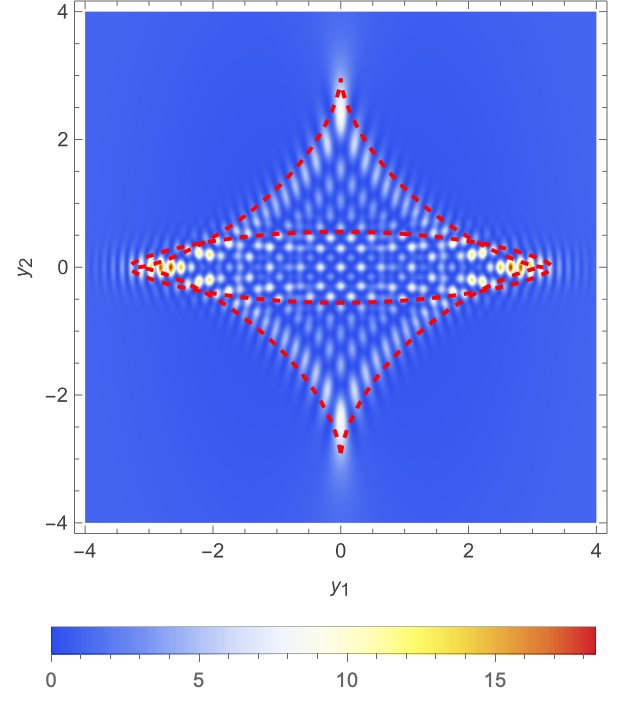}
            \caption{$\omega=10$}
        \end{subfigure}
        \begin{subfigure}[b]{0.32\textwidth}
            \includegraphics[width=\textwidth]{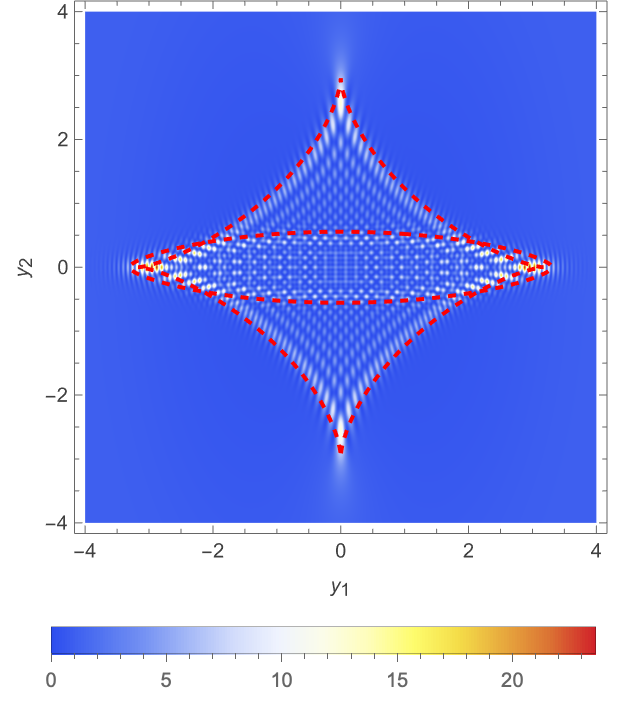}
            \caption{$\omega=20$}
        \end{subfigure}
        \begin{subfigure}[b]{0.32\textwidth}
            \includegraphics[width=\textwidth]{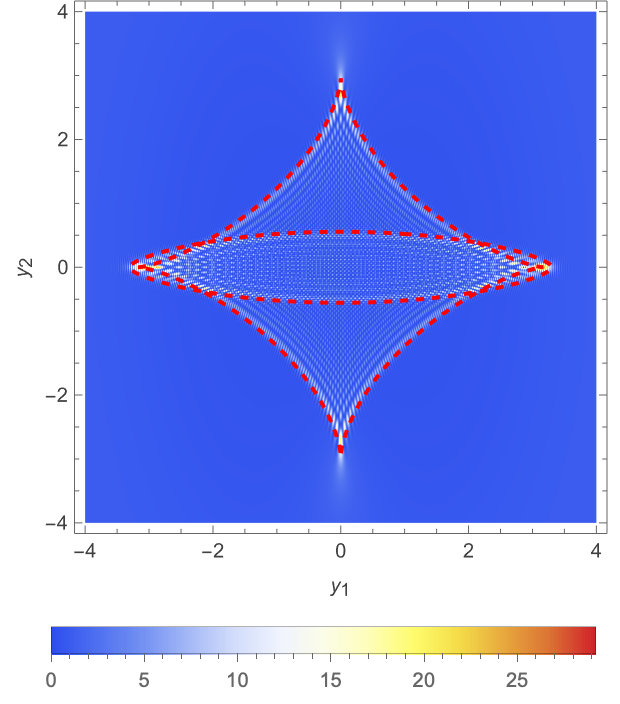}
            \caption{$\omega=50$}
        \end{subfigure}
        \caption{The interference pattern $I(\bm{y}) = |\Psi(\bm{y})|^2$ for the two-dimensional Gaussian lens with the parameters $\alpha  = 7$, $\bm{\mu}=\bm{0}$ and $\Sigma = \text{diag}(1, 2)$ and the associated caustics (the red curves).}\label{fig:Gaussian2D}

        \begin{subfigure}[b]{0.32\textwidth}
            \includegraphics[width=\textwidth]{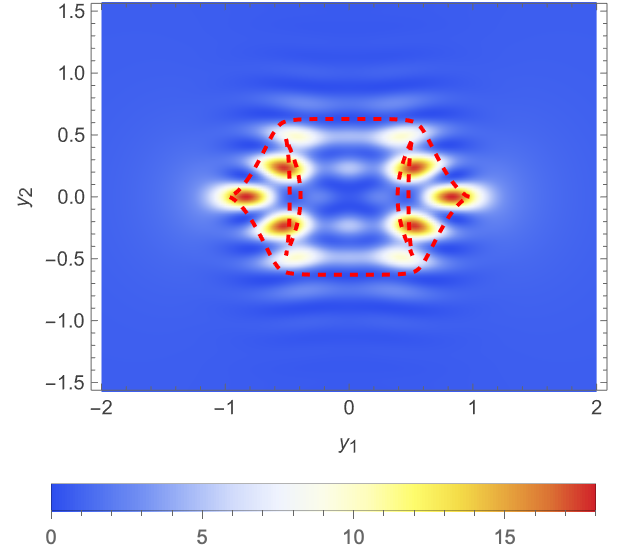}
            \caption{$\omega=10$}
        \end{subfigure}
        \begin{subfigure}[b]{0.32\textwidth}
            \includegraphics[width=\textwidth]{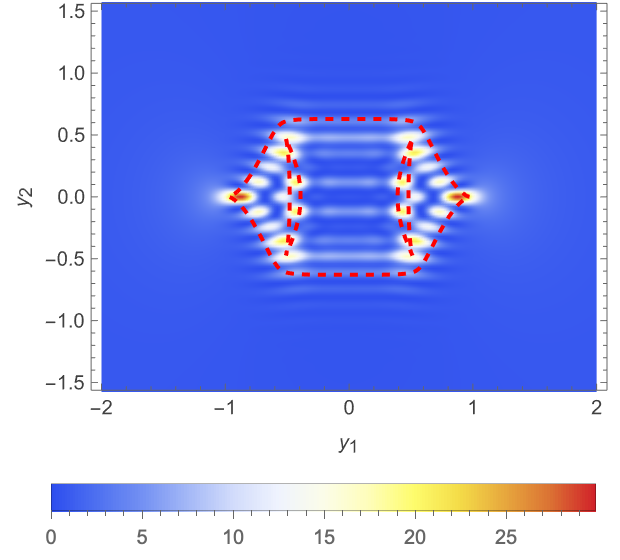}
            \caption{$\omega=20$}
        \end{subfigure}
        \begin{subfigure}[b]{0.32\textwidth}
            \includegraphics[width=\textwidth]{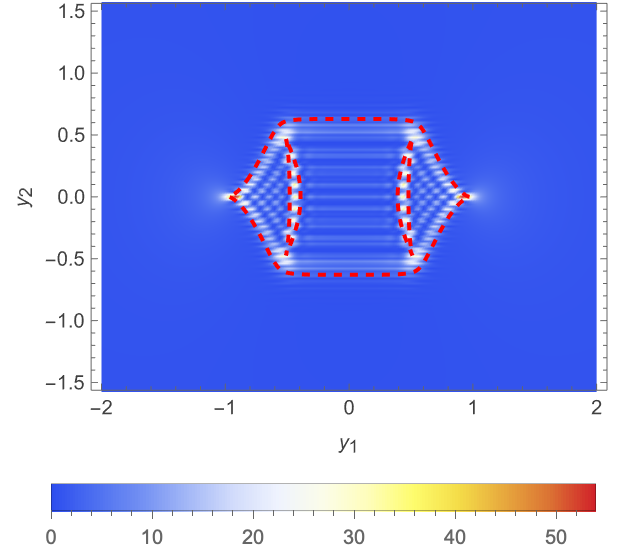}
            \caption{$\omega=50$}
        \end{subfigure}
        \caption{The interference pattern $I(\bm{y}) = |\Psi(\bm{y})|^2$ for the phase-variation consisting of two Gaussian curves $\varphi(\bm{x}) = 2 e^{-\frac{(x_1-1)^2 + x_2^2}{2}} + 2 e^{-\frac{(x_1+1)^2 + x_2^2}{2}}$ with the associated caustics (the red curves).}\label{fig:DoubleGaussianLens2D}
    \end{figure}
    Given the closed-form expression for the coefficients $a_n$, we efficiently evaluate each term to the desired accuracy, ensuring the numerical stability of the diffractive expansion in both the diffractive and refractive regimes. See \cref{fig:Gaussian,fig:Gaussian2D} for the resulting interference patterns of the one- and two-dimensional Gaussian lens in the refractive regime. The lens amplitude converges to the amplitude evaluated with (Picard-Lefshetz numerical methods \cite{Feldbrugge:2023b}).
    
    Next, we consider approximating an arbitrary bounded and integrable one-dimensional lens model by a sum of Gaussian lenses,
    \begin{align}
        \varphi(\bm{x}) \approx \sum_{r=1}^N \alpha_r \varphi_r(\bm{x} - \bm{\mu}_r)\,, \text{ with }\quad \varphi_r(\bm{x}) = e^{-\frac{1}{2} \bm{x}^T\Sigma^{-1}\bm{x}}\,,
    \end{align}
    for a set of amplitudes $\alpha_r$ and positions $\bm{\mu}_r$. For a given phase variation $\varphi$, we can always obtain such an approximation by using radial interpolation to approximate the phase variation by a sum of Gaussian curves \cite{Powell:1977, Wendland:2004}. Using the multinomial theorem
    \begin{align}
        [\varphi_1(\bm{x}) + \dots + \varphi_N(\bm{x})]^n  
        = 
        \sum_{k_1+k_2+\dots+k_N=n} \binom{n}{k_1,k_2,\dots,k_N}\varphi_1(\bm{x})^{k_1}\varphi_2(\bm{x})^{k_2}\dots\varphi_r(\bm{x})^{k_N}\,,
    \end{align}
    with the multinomial $\binom{n}{k_1,\dots, k_N} = \frac{n!}{k_1!\dots k_N!}$, we obtain a diffractive expansion $\sum_{n=0}^\infty a_n(\bm{y}) \omega^n$ with the coefficients
    \begin{align}
        a_n(\bm{y}) 
        &= \left(\frac{\omega}{2 \pi i}\right)^{1/2}\frac{i^n}{n!}  \sum_{k_1+k_2+\dots+k_N=n} \binom{n}{k_1,k_2,\dots,k_N}
        \alpha_1^{k_1}\dots \alpha_N^{k_N} \nonumber\\
        &\times \int e^{i \omega \frac{(\bm{x}-\bm{y})^2}{2}}
            e^{-\frac{1}{2}\sum_{r=1}^N k_r (\bm{x}-\bm{\mu}_r)^T\Sigma^{-1}(\bm{x}-\bm{\mu}_r)}\mathrm{d}\bm{x}\\
        &=\left(\frac{\omega}{2 \pi i}\right)^{1/2} \frac{i^n}{n!}  \sum_{k_1+k_2+\dots+k_N=n} \binom{n}{k_1,k_2,\dots,k_N}
        \alpha_1^{k_1}\dots \alpha_N^{k_N} \int e^{i \omega \frac{(\bm{x}-\bm{y})^2}{2}}
            e^{ \bm{x}^TA\bm{x} + \bm{B} \cdot  \bm{x} +C}\mathrm{d}\bm{x}\\
        &=   \sum_{k_1+k_2+\dots+k_N=n} \frac{i^n \alpha_1^{k_1}\dots \alpha_N^{k_N}}{k_1!k_2!\cdots k_N!}
         \sqrt{\frac{-i \omega}{\det \left(-i \omega I - 2A\right)}}e^{\frac{1}{2} (\bm{B}-i \omega \bm{y})^T(-i \omega I - 2 A)^{-1}(\bm{B} - i \omega \bm{y}) + \frac{i \omega \bm{y}^2}{2}}
    \end{align}
    where the sum is taken over $N$ positive integers $k_r \geq 0$ summing to $n$, with the constants
    
    \begin{align}
            A = -\frac{1}{2}\sum_{r=0}^N k_r \Sigma^{-1}\,, \quad
            \bm{B} = \Sigma^{-1} \left(\sum_{r=1}^N k_r \bm{\mu}_r\right)\,, \quad
            C = -\frac{1}{2}\sum_{r=1}^N k_r \bm{\mu}_r^T \Sigma^{-1} \bm{\mu}_r\,.
    \end{align}
    See \cref{fig:DoubleGaussianLens2D} for an illustration of the lens integral, obtained with this method, for a lens model with a phase-variation consisting of two Gaussian curves.

    The diffractive expansion of the Gaussian lens enables the analytic evaluation of a large class of lens models and does not require any analytic properties of the phase variation models. However, for large frequencies and when summing many Gaussian lenses, this calculation can become computationally expensive. Remarkably, the convergence rate is not significantly influenced by the dimensionality of the integral. As for high-dimensional Gaussian lenses and multidimensional lens problems, determining all the relevant classical rays in the refractive approximation (discussed in \cref{sec:refractiveregime}) can be exponentially hard. The diffractive expansion provides an efficient method to evalaute the lens amplitude for these problems. The implementation of this procedure for multiplane lens systems in wave optics and for Feynman path integrals will be explored in a future paper.

    \section{The refractive expansion}\label{sec:refractiveregime}
    The diffractive expansion, typically used in the refractive regime $\omega \gg 1$, offers a conceptually transparent method for approximating the lens integral that is particularly effective in the low-frequency regime. The refractive expansion, discussed in the present section, provides a complementary perspective that reveals the interplay between real and complex classical rays and their associated caustics. Traditionally, the refractive expansion is understood as the eikonal approximation in the large-frequency limit of the Kirchhoff-Fresnel integral \cite{Schneider:1992}. By employing the theory of resurgence we outline how classical rays and their associated divergent asymptotic series, assembled into a transseries, are intrinsically connected and enable the approximation of the lens integral to arbitrary accuracy at any frequency. In principle, the refractive expansion can even be extended to the diffractive regime!

    We begin with a concise review of the geometric optics approximation, Picard-Lefschetz theory and the eikonal approximation, emphasising their relationship to caustics. We then move beyond the eikonal regime to derive a transseries representation of the lens integral. We develop the finite interpretation of the transseries through the framework of superasymptotic and hyperasymptotic approximations in resurgence theory in \cref{sec:resurgence}.

    \subsection{Classical rays and Picard-Lefschetz theory}
    Lensing is typically introduced through the study of rays in the geometric optics approximation \cite{Schneider:1992, Congdon:2018}. In this picture, radiation behaves like a classical projectile, deflected as it moves through the lens. Yet, remarkably, the concept of rays remains significant well beyond the geometric optics regime.

    \subsubsection{Geometric optics}
\begin{figure}
        \centering
        \begin{subfigure}[b]{0.32\textwidth}
            \includegraphics[width=\textwidth]{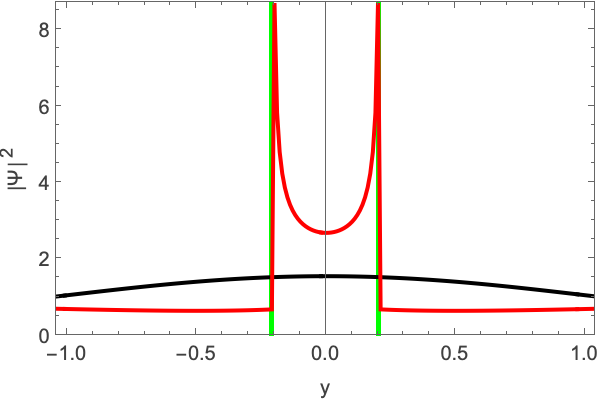}
            \caption{$\omega=1$}
        \end{subfigure}
        \begin{subfigure}[b]{0.32\textwidth}
            \includegraphics[width=\textwidth]{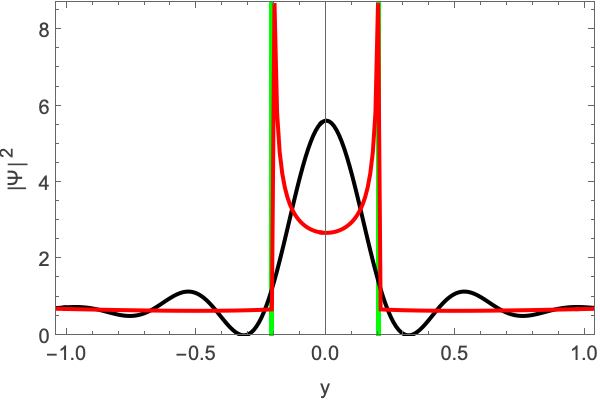}
            \caption{$\omega=10$}
        \end{subfigure}
        \begin{subfigure}[b]{0.32\textwidth}
            \includegraphics[width=\textwidth]{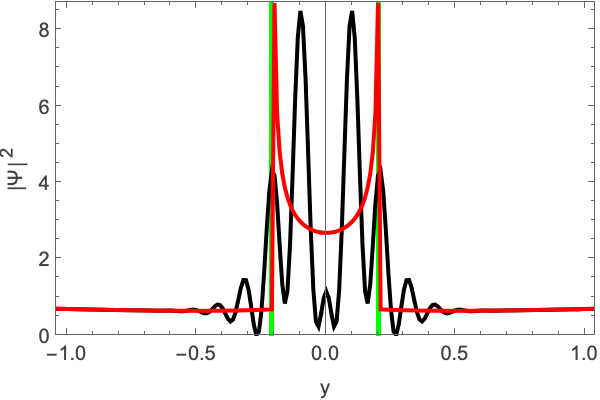}
            \caption{$\omega=50$}
        \end{subfigure}
        \caption{The geometric approximation $I_{\text{Geometric}}$ (red) and the lens integral $|\Psi|^2$ for the one-dimensional Lorentzian lens with amplitude $\alpha=1$ as a function of the frequency $\omega = 1, 10,$ and  $50$.}\label{fig:GO}
    \end{figure}

    When the wavelength of the radiation is much smaller than the characteristic length scales of the lens, or the source emits incoherent radiation, wave effects can be neglected, and the lens can be accurately described within the \textit{geometric optics approximation}. Following Fermat's principle, a classical ray is a stationary point of the time delay function, \textit{i.e.},
    \begin{align}
        \nabla_{\bm{x}}T(\bm{x},\bm{y})&= \bm{x} - \bm{y} + \nabla \varphi(\bm{x}) = \bm{0}\,.
    \end{align}
    Solving for $\bm{y}$, we obtain the lens map
    \begin{align}
        \bm{\xi}(\bm{x}) = \bm{x} + \nabla \varphi (\bm{x})\,.\label{eq:LagrangianMap}
    \end{align}
    A ray intersecting the lens at $\bm{x}$ is deflected by the angle $\nabla \varphi(\bm{x})$ and forms an image at $\bm{y} = \bm{\xi}(\bm{x})$. The intensity follows from the convergence and divergence of a congruence of rays,
    \begin{align}
        I_{\text{geometric}}(\bm{y}) =  \sum_{\bm{x} \in \bm{\xi}^{-1}(\bm{y})} \frac{1}{|\det \nabla \bm{\xi}(\bm{x})|}\,.\label{eq:Geometric_Optics}
    \end{align}
    The intensity receives a contribution from each real classical ray solving the equation $\bm{\xi}(\bm{x})=\bm{y}$. If a bundle of rays around a classical ray spreads (contracts), the determinant of the deformation tensor $\nabla \bm{\xi}$ is larger (smaller) than one and intensity receives a smaller (larger) contribution from this particular ray. When a point in the image plane $\bm{y}$ is reached by $m$ distinct rays -- when $\bm{y}=\bm{\xi}(\bm{x})$ has $m$ real solutions -- it is part of an $m$-image region. See \cref{fig:GO} for an illustration of the geometric optics approximation of the one-dimensional Lorentzian lens. The geometric optics approximation spikes in two, so-called, fold caustics separating two one-image regions from a three-image region in the middle.
    
    The intensity spikes to infinity when the determinant of the deformation tensor $\nabla \bm{\xi} = I + \mathcal{H} \varphi$ vanishes, with the Hessian of the phase-variation $\mathcal{H} \varphi$. Such a point  $\bm{y}_c$ corresponds to a degenerate critical point $\bm{x}_c$ of the time delay for which both the gradient $\nabla_{\bm{x}} T(\bm{x}_c,\bm{y}_c)$ and the Hessian $\mathcal{H}_{\bm{x}}T(\bm{x}_c,\bm{y}_c)$ is singular. Caustics come in a variety of morphologies that are famously classified by catastrophe theory in terms of the elementary catastrophes \cite{Thom:1974, Arnold:1985,Berry:1980,Nye:1999}. An excellent and concise pedagogical introduction to catastrophe theory and its applications in the physical sciences can be found in the textbook \cite{Saunders:1980}. The $m$-image regions are bounded by the fold caustic ($A_2$), defined by the eigenvalue field of the deformation tensor
    \begin{align}
        \lambda_i(\bm{x}) = 0\,,
    \end{align}
    for some $i \in \{1,2,\dots,d\}$, where the eigenvalue $\lambda_i$ and the eigenvector $\bm{v}_i$ of the deformation tensor are defined by the eigenequation
    \begin{align}
        \nabla \bm{\xi}(\bm{x}) \bm{v}_i(\bm{x}) = \lambda_i(\bm{x}) \bm{v}_i(\bm{x})\,.
    \end{align}
    In terms of the eigenvalue fields, the intensity is then given by
    \begin{align}
        I_{\text{geometric}}(\bm{y}) =  \sum_{\bm{x} \in \bm{\xi}^{-1}(\bm{y})} \frac{1}{|\lambda_1(\bm{x})|\cdots|\lambda_d(\bm{x})|}\,.
    \end{align}
    In geometric optics, the fold caustics mark spikes in the intensity pattern (see, for example, the green vertical lines in \cref{fig:GO}). In wave optics, the fold caustics bound the interference patterns (see, for example, the red curves in \cref{fig:Gaussian2D,fig:DoubleGaussianLens2D}). When, in addition, the fold curve is parallel to the eigenvector field,
    \begin{align}
        \bm{v}_i(\bm{x}) \cdot \nabla \lambda_i(\bm{x}) = 0\,,
    \end{align}
    the degenerate singularity forms a cusp caustic (for a derivation see, for example, \cite{Feldbrugge:2018}). The cusp ($A_3$) caustics emerge as kinks in the fold caustic (as can be seen in \cref{fig:Gaussian2D,fig:DoubleGaussianLens2D}). The fold and the cusp are the only caustics that form in generic two-dimensional lens problems. Oscillatory integrals with more external parameters can exhibit more complicated caustics, including the swallowtail ($A_4$), the butterfly ($A_5$), the elliptic ($D_4^+$), the hyperbolic ($D_4^-$) and the parabolic ($D_5$) caustics. For a detailed review of catastrophe theory and caustics in optics, we refer to \cite{Berry:1980,Nye:1999}. For an analysis of caustics in radio astronomy, see \cite{Melrose}.

    \subsubsection{Picard-Lefschetz theory}\label{sec:PL}
\begin{figure}
        \centering
        \begin{subfigure}[b]{0.32\textwidth}
            \includegraphics[width=\textwidth]{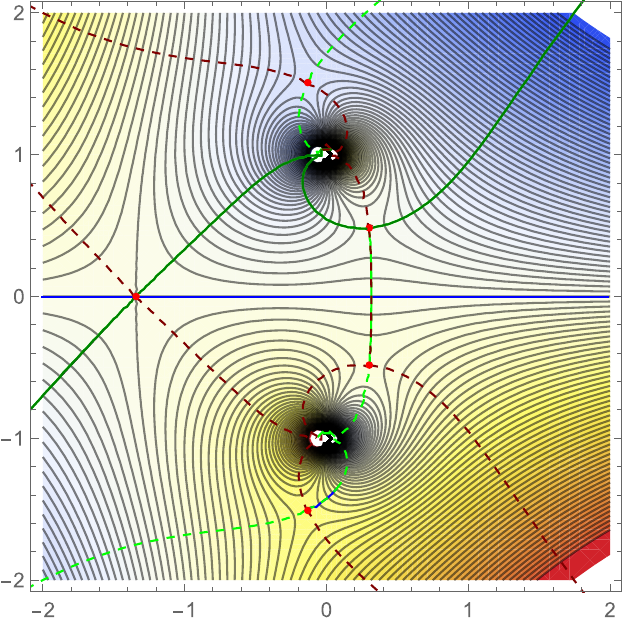}
            \caption{$y=-1$}
        \end{subfigure}
        \begin{subfigure}[b]{0.32\textwidth}
            \includegraphics[width=\textwidth]{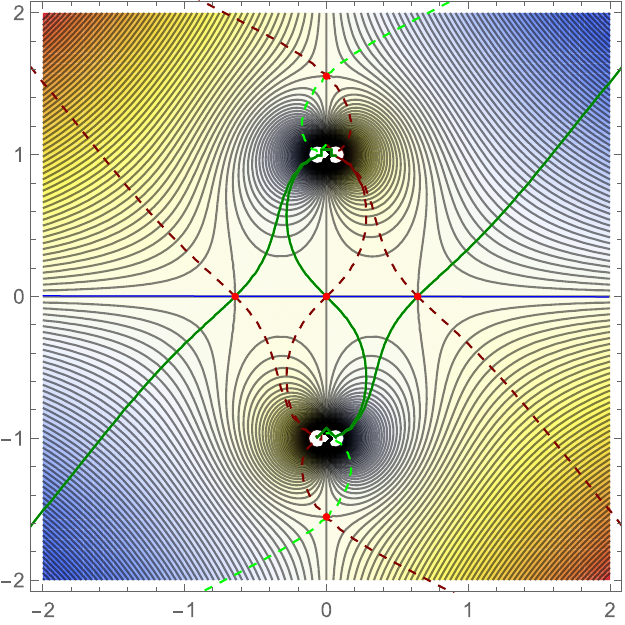}
            \caption{$y=0$}
        \end{subfigure}
        \begin{subfigure}[b]{0.32\textwidth}
            \includegraphics[width=\textwidth]{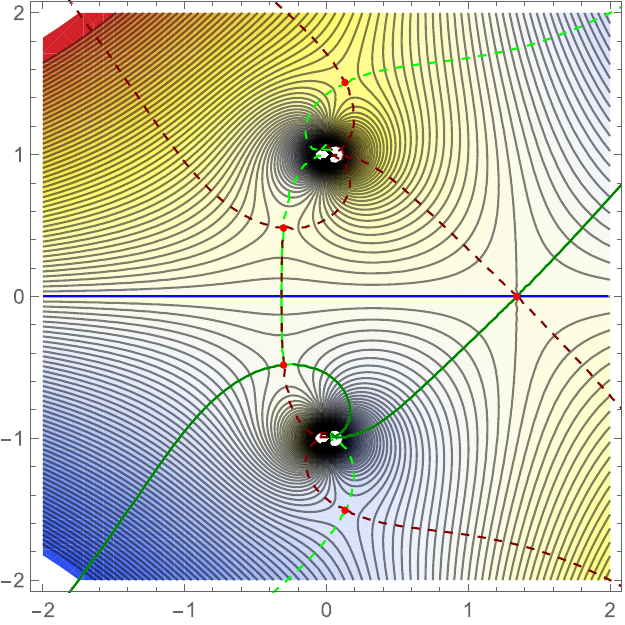}
            \caption{$y=1$}
        \end{subfigure}
        \caption{Optimal deformation of the integration domain in the complex $x$-plane for the one-dimensional Lorentzian lens with amplitude $\alpha=1$.}\label{fig:PL_deformation}
        \begin{subfigure}[b]{0.32\textwidth}
            \includegraphics[width=\textwidth]{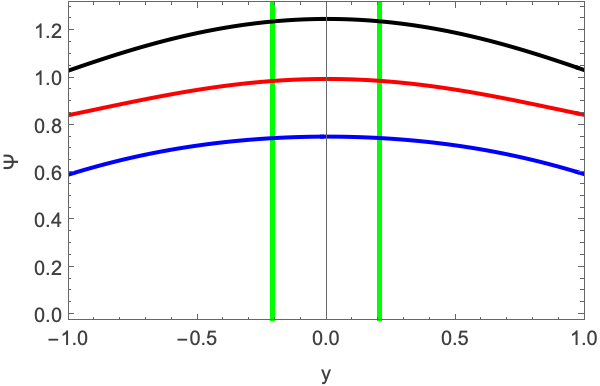}
            \caption{$\omega=1$}
        \end{subfigure}
        \begin{subfigure}[b]{0.32\textwidth}
            \includegraphics[width=\textwidth]{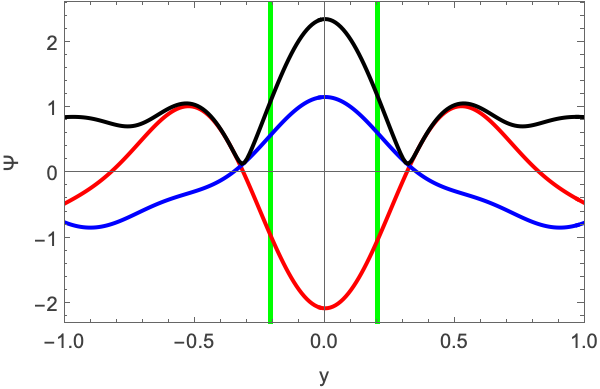}
            \caption{$\omega=10$}
        \end{subfigure}
        \begin{subfigure}[b]{0.32\textwidth}
            \includegraphics[width=\textwidth]{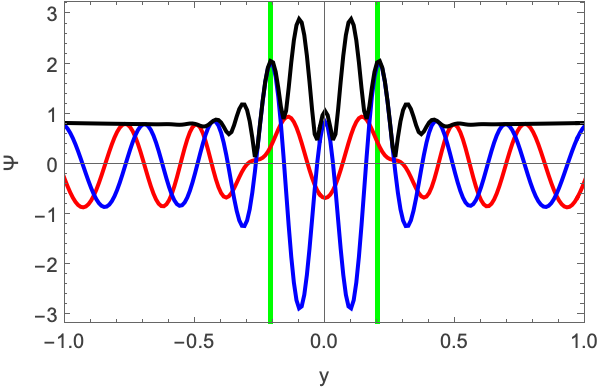}
            \caption{$\omega=50$}
        \end{subfigure}
        \caption{The diffraction integral of the one-dimensional Lorentzian lens with amplitude $\alpha=1$ evaluated with Picard-Lefschetz theory. The red, blue and black curves correspond to the real part, the imaginary part and the modulus of $\Psi(y)$ as a function of the frequency $\omega=1, 10,$ and $50$. The red lines mark the caustics.}\label{fig:PL}
    \end{figure}

    The classical rays not only govern the geometric optics approximation but are also key to understanding diffractive expansions of the lens amplitude through \textit{Picard-Lefschetz theory} \cite{Feldbrugge:2017,Feldbrugge:2023b}. The first introduction of Picard-Lefschetz theory to theoretical physics is found in \cite{Witten:2010} and for a pedagogical introduction and the first application to quantum cosmology we refer the reader to \cite{Feldbrugge:2017}. The Kirchhoff-Fresnel integral \cref{eq:KF} is highly oscillatory and conditionally convergent, making the definition delicate and the direct numerical evaluation expensive \cite{Feldbrugge:2023}. Assuming the phase variation is a meromorphic function, we may analytically continue the time delay into the complex plane and deform the integration domain to improve the convergence properties of the integral. Cauchy's integral theorem guarantees that the lens amplitude is preserved as long as we keep the endpoints (boundary of the integration domain) fixed, and we do not deform the integration contour past a singularity of the integrand. Writing the exponent in terms of a real and an imaginary part, 
    \begin{align}
        i T(\bm{x}, \bm{y}) = h(\bm{x},\bm{y}) + i H(\bm{x},\bm{y})\,,
    \end{align}
    we note that the real part $h = \log |e^{i \omega T}|$ governs the amplitude while the imaginary part $H = \text{arg}(e^{i \omega T})$ governs the oscillatory nature of the integrand. By flowing the original integration domain along the downward flow of the real part $h$,
    \begin{align}
        \frac{\mathrm{d} \bm{\gamma}_\lambda(\bm{x}_0)}{\mathrm{d} \lambda} = - \nabla h(\bm{\gamma}_\lambda(\bm{x}_0))\,,
    \end{align}
    starting in $\bm{\gamma}_{\lambda=0}(\bm{x}_0) = \bm{x}_0$ and defining the gradient as one would on $\mathbb{C}^d \simeq \mathbb{R}^{2d}$, the integration domain is deformed into the complex plane. Such a deformation preserves the value of the integral while improving the convergence properties of the integrand. In the limit $\lambda \to \infty$, the original integration domain transforms into a sum of steepest descent manifolds\footnote{
        The descent manifold $\mathcal{J}_j$ of a classical ray $\bm{x}_j$ consists of the points $\bm{x}_0$ for which the downward flow $\bm{\gamma}_\lambda(\bm{x}_0)$ reaches the classical ray $\bm{x}_j$ in the limit of $\lambda \to -\infty$.} $\sum_j n_j \mathcal{J}_j$ associated to a set of real and complex classical rays $\bm{x}_j$ -- complex solution of of $\bm{\xi}(\bm{x})=\bm{y}$ -- forming the Picard-Lefschetz formula  
    \begin{align}
        \Psi(\bm{y}) 
        &= \left(\frac{\omega}{2 \pi i}\right)^{d/2} \sum_j n_j(\bm{y}) \int_{\mathcal{J}_j} e^{i \omega T(\bm{x}, \bm{y})}\mathrm{d}\bm{x}\\
        &= \left(\frac{\omega}{2 \pi i}\right)^{d/2} \sum_j n_j(\bm{y})  e^{i \omega H(\bm{x}_j, \bm{y})}\int_{\mathcal{J}_j} e^{\omega h(\bm{x}, \bm{y})}\mathrm{d}\bm{x}\,.\label{eq:PL}
    \end{align} 
    The inclusion of complex solutions to the lens equation $\bm{\xi}(\bm{x})= \bm{y}$ is central to wave optics! The key insight of Picard and Lefschetz is that the steepest descent manifold $\mathcal{J}_j$, associated with the classical ray $\bm{x}_j$, is part of the deformation if and only if the associated steepest ascent manifold\footnote{
        The ascent manifold $\mathcal{K}_j$ of a classical ray $\bm{x}_j$ consists of the points $\bm{x}_0$ for which the downward flow $\bm{\gamma}_\lambda(\bm{x}_0)$ reaches the classical ray $\bm{x}_j$ in the limit $\lambda \to +\infty$.} 
    $\mathcal{K}_j$ intersects the original integration domain, $n_j = \langle \mathbb{R}^d, \mathcal{K}_j \rangle  \in \mathbb{Z}$ (the bracket $\langle J, K \rangle$ vanishes when the manifolds $J$ and $K$ do not intersect). When the steepest descent manifold of a classical ray $\bm{x}_j$ is included, the classical ray is said to be \textit{relevant}. When it is not part of the deformation, the classical ray is \textit{irrelevant}. Note that a real classical ray is always relevant. The intersection number of a complex ray can be evaluated by studying the descent flow (for numerical implementations see \cite{Feldbrugge:2023b, Shoji:2025} and \cite{Weber:2025}). We illustrate the Picard-Lefschetz deformation for the one-dimensional Lorentzian lens in \cref{fig:PL_deformation}. As the imaginary part $H$ of the integrand is preserved by the downward flow,\footnote{
        Writing the point on the lens plane and the path $\bm{\gamma}_\lambda$ in terms of a real and an imaginary parts $\bm{x} = \bm{u} + i \bm{v}$ and $\bm{\gamma}_\lambda=\bm{u}_\lambda + i \bm{v}_\lambda$, satisfying the flow equations $\frac{\mathrm{d} \bm{u}_\lambda}{\mathrm{d} \lambda} = - \nabla_{\bm{u}} h$ and $\frac{\mathrm{d} \bm{v}_\lambda}{\mathrm{d} \lambda} = - \nabla_{\bm{v}} h$ (with $\nabla_{\bm{u}}$ and $\nabla_{\bm{v}}$ the gradient in the real and imaginary directions), we find that
    \begin{align}
        \frac{\mathrm{d} H(\bm{\gamma}_\lambda(\bm{x}_0))}{\mathrm{d} \lambda} 
        &= 
        \nabla_{\bm{u}} H \cdot \frac{\mathrm{d} \bm{u}_\lambda}{\mathrm{d}\lambda}
        +
        \nabla_{\bm{v}} H \cdot \frac{\mathrm{d} \bm{v}_\lambda}{\mathrm{d}\lambda}\\
        &= 
        -\nabla_{\bm{u}} H \cdot \nabla_{\bm{u}}h
        -
        \nabla_{\bm{v}} H \cdot \nabla_{\bm{v}}h\\
        &= - \nabla h \cdot \nabla H\\
        &= 0\,,
    \end{align}
    by the multidimensional Cauchy-Riemann equations $\nabla h \cdot \nabla H =0$.} the Picard-Lefschetz deformation resums the oscillations or the original integral and renders it absolutely convergent. In fact, one may argue that the original conditionally convergent lens integral is ill-defined -- its value depends on the chosen regulator -- and that the deformed integral \cref{eq:PL} constructed using analyticity is the proper rigorous definition (see \cite{Feldbrugge:2023} for a more detailed discussion). More practically, the Picard-Lefschetz deformation is the optimal deformation (among the contours equivalent to the original integration domain), cancelling all oscillations, and enables the efficient numerical evaluation of the Kirchhoff-Fresnel integral \cite{Feldbrugge:2023b}. See \cref{fig:PL} for the one-dimensional Lorentzian lens amplitude evaluated with the Picard-Lefschetz numerical integration scheme. Below, we verify the accuracy of the diffractive approximations by comparing the approximation with the numerically evaluated lens amplitude.

\begin{figure}
        \centering
        \begin{subfigure}[b]{0.32\textwidth}
            \includegraphics[width=\textwidth]{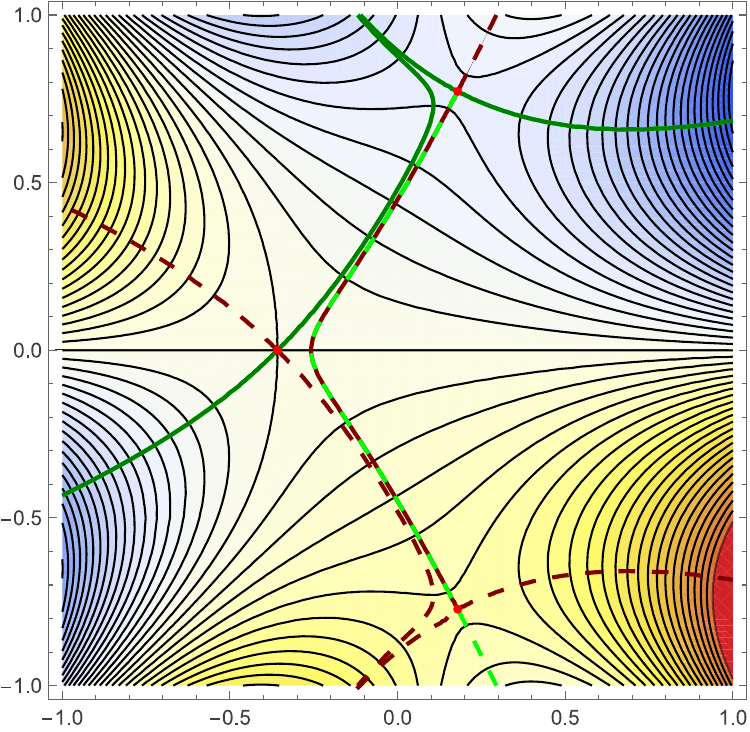}
        \end{subfigure}
        \begin{subfigure}[b]{0.32\textwidth}
            \includegraphics[width=\textwidth]{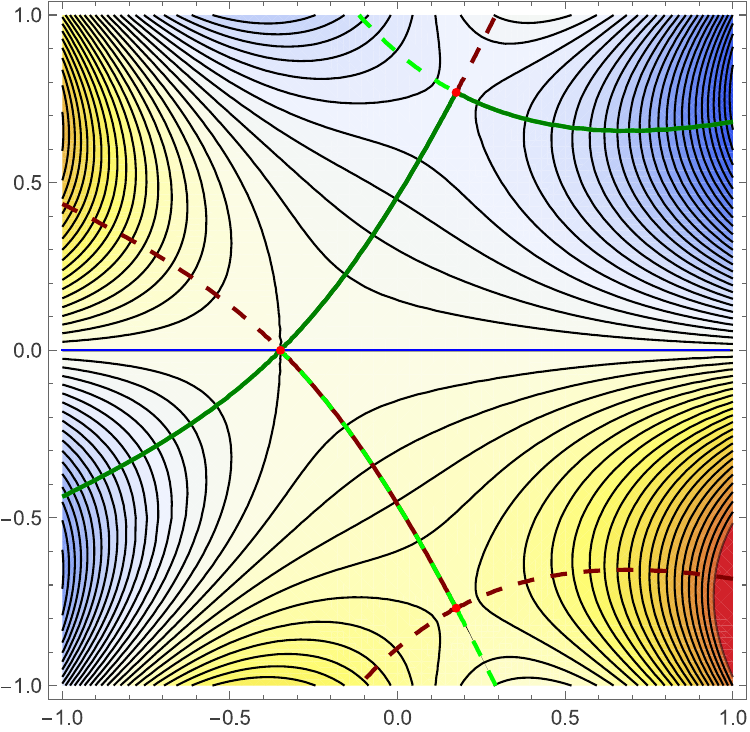}
        \end{subfigure}
        \begin{subfigure}[b]{0.32\textwidth}
            \includegraphics[width=\textwidth]{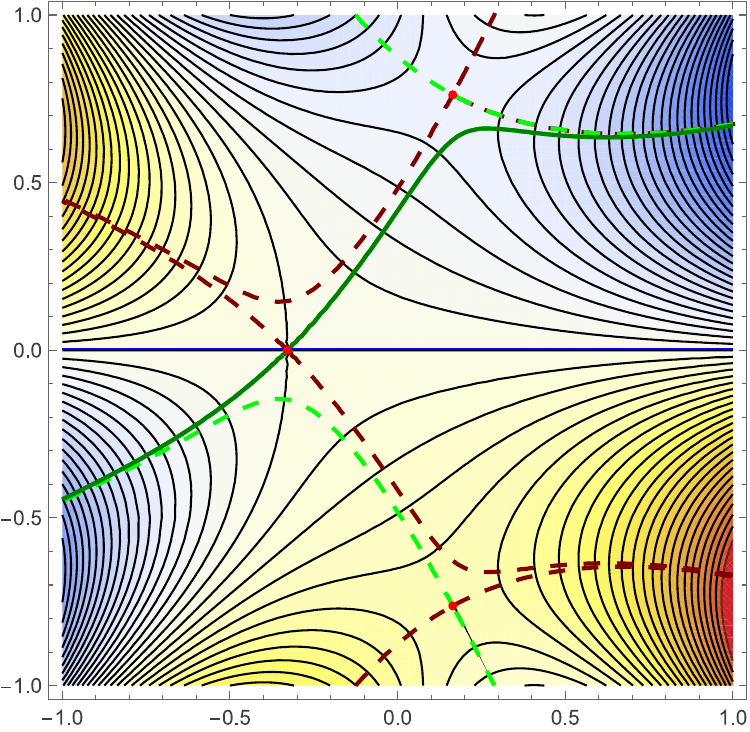}
        \end{subfigure}
        \caption{In a Stokes transition (the central panel), the steepest descent of one of the complex rays terminates on another saddle point. To the left and right of the Stokes transition, the complex saddle point transitions from relevant (the left panel) to irrelevant (the right panel).}\label{fig:Stokes}
    \end{figure}

\begin{figure}
        \centering
        \includegraphics[width=0.5\textwidth]{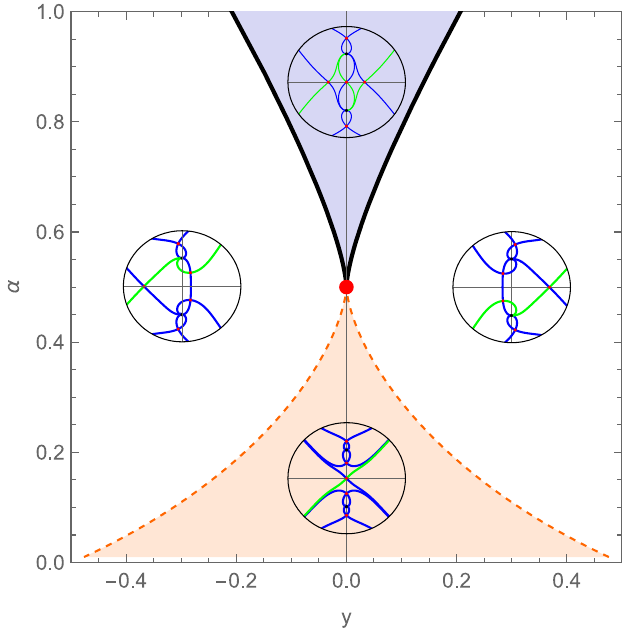}
        \caption{Caustics (black) with the cusp (red) and Stokes lines (blue) in the $y$-$\alpha$ plane.}\label{fig:caustics}
    \end{figure}

    The Picard-Lefschetz deformation changes as a function of $\bm{y}$. At a generic point $\bm{y}$, the change is smooth. However, at a caustic and a Stokes transsition, the nature of the Picard-Lefschetz contour changes dramatically. At the caustics, where the time delay has a degenerate critical point and two classical rays coalesce, the associated steepest descent manifolds merge. After moving through the caustic, two of the coalescing rays become complex classical rays forming a complex conjugate pair. One of these complex classical rays typically remains relevant (the one for which $h$ is negative) while the complex conjugate ray is irrelevant to the integral. At a Stokes transition, where the steepest descent manifold of one classical ray $\bm{x}_j$ terminates at another classical ray $\bm{x}_k$, the relevance of the classical ray $\bm{x}_k$ changes (see \cref{fig:Stokes}). As the gradient descent preserves the $H$-function, a Stokes transition where classical ray $\bm{x}_k$ changes relevance is marked by $H(\bm{x}_j)=H(\bm{x}_k)$ and $h(\bm{x}_j) > h(\bm{x}_k)$. In \cref{fig:caustics}, we outline the Picard-Lefschetz deformation for the Gaussian lens as a function of the position on the image plane $\bm{y}$ and the lens strength $\alpha$. For a more detailed discussion on Picard-Lefschetz theory in wave optics and the efficient numerical evaluation of oscillatory integrals in the complex plane, we refer to \cite{Feldbrugge:2023b}.

    \subsubsection{The eikonal approximation}
\begin{figure}
        \centering
        \begin{subfigure}[b]{0.32\textwidth}
            \includegraphics[width=\textwidth]{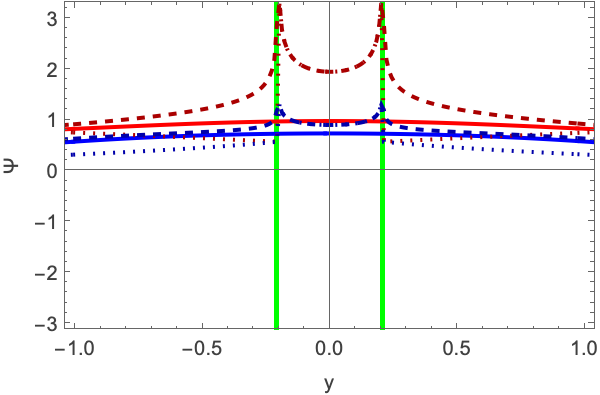}
            \caption{$\omega=1$}
        \end{subfigure}
        \begin{subfigure}[b]{0.32\textwidth}
            \includegraphics[width=\textwidth]{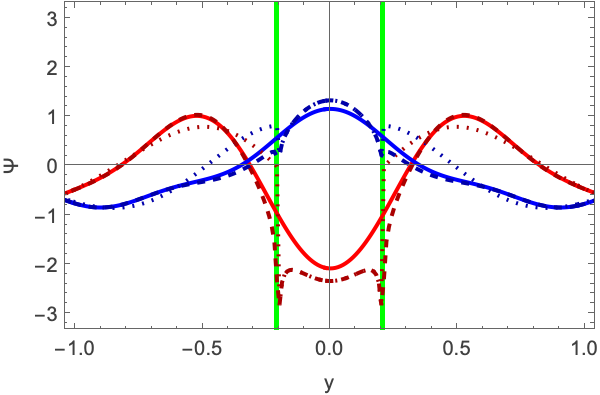}
            \caption{$\omega=10$}
        \end{subfigure}
        \begin{subfigure}[b]{0.32\textwidth}
            \includegraphics[width=\textwidth]{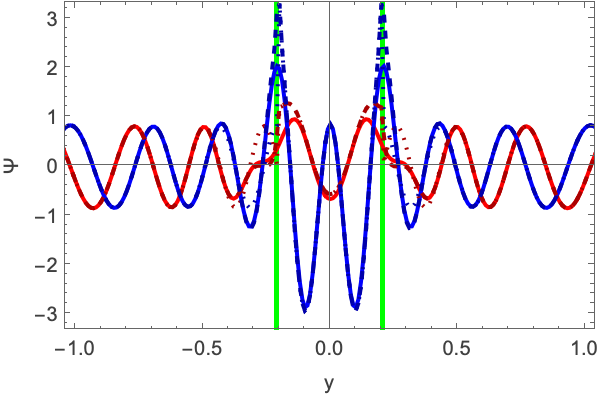}
            \caption{$\omega=50$}
        \end{subfigure}
        \caption{The real and complex eikonal approximation $\Psi_{\text{eikonal}}^\mathbb{R}$ (dotted) and $\Psi_{\text{eikonal}}$ (dashed) compared to the exact result (solid) for the one-dimensional Lorentzian lens with amplitude $\alpha=1$. The red and blue curves correspond to the real and imaginary parts of the lens amplitude. The green lines mark the caustics.}\label{fig:Eikonal}
        \begin{subfigure}[b]{0.32\textwidth}
            \includegraphics[width=\textwidth]{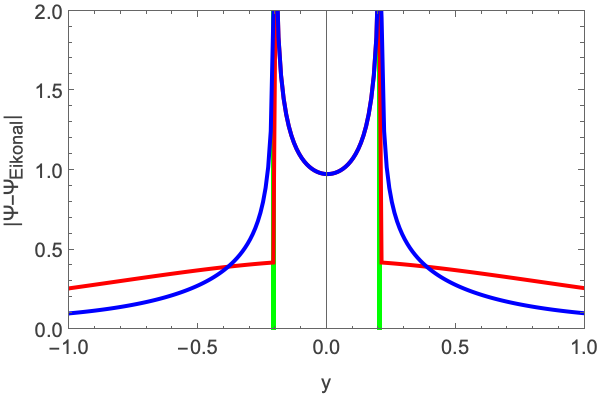}
            \caption{$\omega=1$}
        \end{subfigure}
        \begin{subfigure}[b]{0.32\textwidth}
            \includegraphics[width=\textwidth]{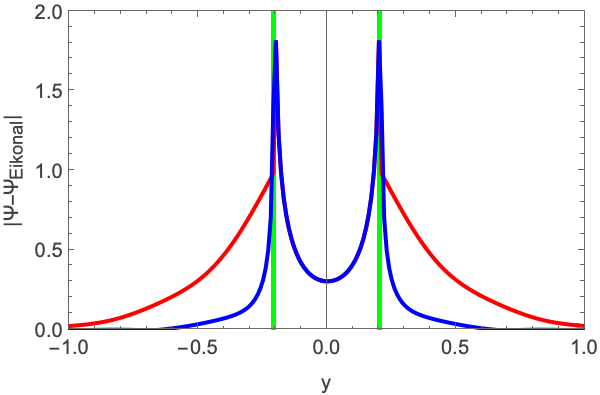}
            \caption{$\omega=10$}
        \end{subfigure}
        \begin{subfigure}[b]{0.32\textwidth}
            \includegraphics[width=\textwidth]{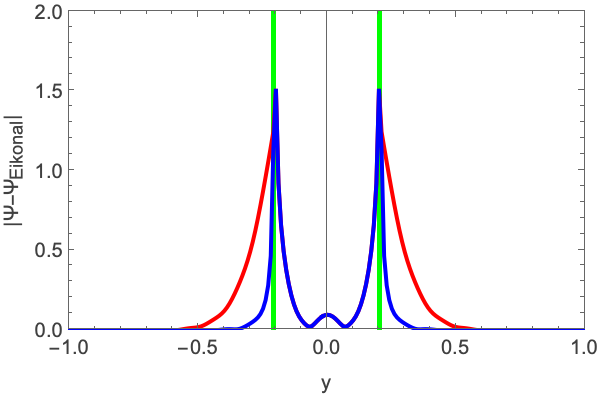}
            \caption{$\omega=50$}
        \end{subfigure}
        \caption{The difference between the eikonal approximation and the exact result $|\Psi(y) - \Psi_{\text{eikonal}}^\mathbb{R}(y)|$ (red) and $|\Psi(y) - \Psi_{\text{eikonal}
        }(y)|$ (blue) for the one-dimensional Lorentzian lens with amplitude $\alpha =1$. The green lines mark the caustics.}\label{fig:Eikonal_error}
    \end{figure}

    We now turn to a brief review of the eikonal approximation. Upon expanding the time delay function in the Picard-Lefschetz formula for the integral over $\mathcal{J}_j$ to quadratic order around the relevant classical rays $\bm{x}_j$, 
    \begin{align}
        T(\bm{x},\bm{y}) \approx T(\bm{x}_j, \bm{y}) + \frac{1}{2} (\bm{x} - \bm{x}_j)^T [\mathcal{H} T(\bm{x}_j,\bm{y})](\bm{x} - \bm{x}_j)\,,
    \end{align}
    the lens integral reduces to a set of Gaussian integrals. Evaluating the Gaussian integrals leads to the eikonal approximation
    \begin{align}\label{eq:complexeikonal}
        \Psi_{\text{eikonal}}(\bm{y}) = \sum_j n_j(\bm{y}) \frac{e^{i \omega T(\bm{x}_j,\bm{y}) }}{\sqrt{\det \nabla \bm{\xi}(\bm{x}_j)}}\,.
    \end{align}
   The eikonal approximation inherits the relevance coefficient $n_j$ from the Picard-Lefschetz deformation discussed in the previous section. The relevant branch of the square root in the denominator follows from continuity (starting, for example, from the free theory $\varphi(\bm{x})=0$). Historically, the eikonal approximation is occasionally defined to only include real classical rays, which are, by construction, all relevant\,, \textit{i.e.},
   \begin{align}
        \Psi_{\text{eikonal}}^{\mathbb{R}}(\bm{y}) = \sum_{\bm{x} \in \bm{\xi}^{-1}(\bm{y})} \frac{e^{i \omega T(\bm{x},\bm{y}) - i m(\bm{x}) \pi / 2}}{\sqrt{|\det \nabla \bm{\xi}(\bm{x})|}}\,.
    \end{align}
    In this real version of the eikonal approximation Riemann sheet structure of the square root is captured by the Morse-index $m$, given by the index of the classical ray. For two-dimensional lenses, the Morse-index is $0,1,$ and $2$ when the ray $\bm{x}_j$ is a minimum, a saddle point and a maximum of the time delay function $T$ respectively. For more details on the eikonal approximation, we refer to \cite{Schneider:1992, Nye:1999}. For one of the first numerical studies of the eikonal approximation of the Kichhoff-Fresnel integral and its relation to caustics, we refer to \cite{Melrose}. This approximation is arguably closer to the geometric optics approximation since squaring the amplitude $|\Psi_{\text{eikonal}}^{\mathbb{R}}|^2$ and ignoring the cross terms we recover the geometric approximation  \eqref{eq:Geometric_Optics}. In \cref{fig:Eikonal,fig:Eikonal_error} we compare the full and real Eikonal approximation with the numerically evaluated Kirchhoff-Fresnel integral for the one-dimensional Lorentzian lens. The two approximations agree away from the caustics where the contributions of complex rays are small. Both approximations diverge near the caustics, where $\det \nabla \bm{\xi} = 0$.
    
    Starting from the geometric optics approximation, the real eikonal approximation is the first step towards the wave nature in lensing. The full eikonal approximation, including relevant complex rays, is a significant improvement in the vicinity of caustics. In \cref{sec:transseries}, we demonstrate how to go beyond the eikonal approximation using the tools of resurgence theory: uniform asymptotics, superasymptotics and hyperasymptotic expansions.
   
    \subsection{Laplace integrals}\label{sec:Laplace}
    In this section we consider the Picard-Lefschetz formula for the Kirchhoff-Fresnel integral from a different perspective. Let us, for convenience, restrict the present discussion to the one-dimensional lens. At the end of this discussion, we generalize to multidimensional integrals. Changing variables, $t = -i T(x,y)$, the lens integral may be interpreted as a Laplace type integral
    \begin{align}
        \Psi(y) = \sqrt{\frac{\omega}{2 \pi i}} \int_{C_y} e^{- \omega t} \phi_y(t)\mathrm{d}t\,,
    \end{align}
    of the function 
    \begin{align}
        \phi_y(t) = \left[\frac{\mathrm{d}t}{\mathrm{d}x}\right]^{-1}\,,
    \end{align} 
    evaluated over the integration domain
    \begin{align}
        C_y = \{ -i T(x,y) \,| x \in \mathbb{R}\}\,.
    \end{align}
    Note that the Jacobian $\phi_y$ is obtained by solving the position in the lens plane $x$ as a function of $t$. This equation has generally multiple solutions, making the function $\phi_y$ multivalued. 

    As will be clarified in \cref{sec:borel}, the complex $t$-plane is known as the Borel plane. Each saddle point $x_j$ of the original integral is transformed into a branch point singularity of the integrand $\phi_y$ at the point $t_j = -i T(x_j)$ in the Borel plane.\footnote{
        A nondegenerate critical point leads to a branch point of the form $1/\sqrt{t}$, as can be seen from the Gaussian integral
        \begin{align}
            \int_{-\infty}^\infty e^{i \omega x^2}\mathrm{d}x = \int e^{-\omega t} \frac{\mathrm{d}x}{\mathrm{d}t}\mathrm{d}t = \int_{C} e^{-\omega t} \frac{1+i}{2 \sqrt{2t}}\mathrm{d}t
        \end{align}
        with the contour $C$ in the Borel plane, running along the imaginary axes from $+ \infty$ to the branch point at $t=0$ back to $+\infty$ at the other Riemann sheet of the square root. Degenerate saddle points still map to singularity branch points. However, the Riemann sheet structure at the branch point changes, as can be observed from the integral
        \begin{align}
            \int_{-\infty}^\infty e^{i \omega x^n}\mathrm{d}x = 
            \int_{C} e^{-\omega t} \frac{e^{\frac{i \pi}{2n}}}{n t^{1-\frac{1}{n}}}\mathrm{d}t\,.
        \end{align}
        }
    The multivalued nature of the Riemann surface $\phi_y$ is a central feature of the lens integral considered as a Laplace-type integral. The integration contour $C_y$ should be interpreted as a curve on the Riemann sheets of the Borel transform $\phi_y$, navigating the branch points. See \cref{ap:algebraic_curve} for a more detailed discussion on this algebraic curve/Riemann surface.

\begin{figure}
        \centering
        \begin{subfigure}[b]{0.45\textwidth}
            \includegraphics[width=\textwidth]{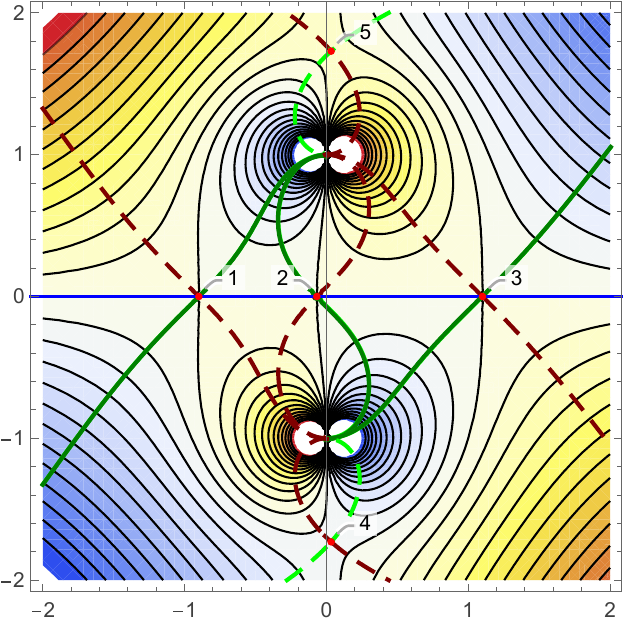}
        \end{subfigure}
        \begin{subfigure}[b]{0.45\textwidth}
            \includegraphics[width=\textwidth]{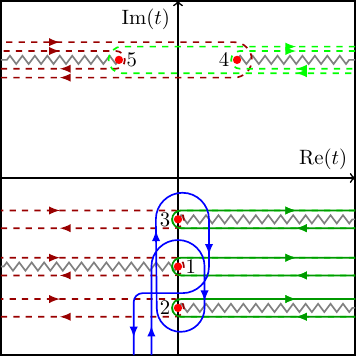}
        \end{subfigure}
        \caption{The Picard-Lefschetz deformation for the one-dimensional Lorentzian lens with the parameters $\alpha=2$, $y=1/2$. \textit{Left:} The complex $x$-plane with the classical rays points $x_j$ (the labelled red points), the associated steepest descent $\mathcal{J}_j$ (the green curves) and ascent manifolds $\mathcal{K}_j$ (the red curves), and the optimal deformation of the real line (the dark green dashed curves). \textit{Right:} The associated branch points $t_j$ (the labelled red points), the steepest descent $\mathcal{BJ}_j$ (the green curves) and ascent manifolds $\mathcal{BK}_j$ (the red curves) in the Borel plane around the branch cuts (the grey zigzag lines). The original integration domain is mapped to the blue contour $C_y$ navigating the branch cuts.}\label{fig:LorentzianBorel}
    \end{figure}

    The Laplace integral provides a powerful perspective on the Picard-Lefschetz deformation, as the descent flow preserves the imaginary part of the exponent. More specifically, the original integration domain, the steepest ascent and descent contours, transform into line segments $C_y$, $\mathcal{BJ}_j$ and $\mathcal{BK}_j$ in the Borel plane. 
    The contour $C_y$ is a curve along the line $\text{Re}[t] = 0$ looping around the branch points associated with the real classical rays. The steepest descent $\mathcal{B}\mathcal{J}_j$ and steepest ascent contours $\mathcal{B}\mathcal{K}_j$ in the Borel plane form horizontal lines, $\text{Im}[t]=\text{const}$, on the Riemann sheet. For an illustration, we consider the Picard-Lefschetz deformation of the Lorentzian lens in both the complex $x$-plane and the Borel plane (see \cref{fig:LorentzianBorel}). We observe that the steepest ascent and descent manifolds simplify, while the original integration domain $C_y$ loops around the branch cuts (flowing around $t_1$, $t_2$ and finally $t_3$). The relevance of a saddle point can be determined by intersecting $C_y$ with $\mathcal{BK}_j$. A Stokes phenomenon corresponds to the configuration where two branch points $t_j$ and $t_k$ have the same imaginary part and ``see" each other on the Riemann sheet in terms of the radar method of Voros \cite{Voros:1983} and Berry and Howls \cite{Berry:1991}. At a Stokes transition, one branch point overtakes another branch point in the Borel plane, \textit{i.e.}, a Stokes transition occurs when the branch points are on the same horizontal line $\text{Im } t_j = \text{Im } t_k$ and are adjacent on the Riemann sheet.
    
    The Picard-Lefschetz formula leads to the expression
    \begin{align}
        \Psi(y) = \sqrt{\frac{\omega}{2\pi i}} \sum_j n_j(y) e^{i \omega T(x_j, y)} \int_0^\infty e^{- \omega t} \Delta \phi_y^{(j)}(t)\mathrm{d}t,
    \end{align}
    with $\Delta \phi_y^{(j)}$ the difference of $\phi_j$ on the two Riemann sheets evaluated along the associated horizontal branch cuts in the Borel plane. As we will see in \cref{sec:borel}, the function $\Delta \phi_y$ is known as the Borel transform.  

    \bigskip
    For multidimensional integrals, the Borel transform is obtained by integrating over the level sets of the time delay restricted to a steepest descent manifold $\mathcal{J}_j$,
    \begin{align}
        \Delta \phi_{\bm{y}}^{(j)}(t) =  \int_{\{\bm{x} \in \mathcal{J}_j\,|\, T(\bm{x},\bm{y})=t\}} \frac{\mathrm{d} \bm{x}}{\|\nabla_{\bm{x}} h(\bm{x},\bm{y})\|}\,.
    \end{align}
    The multidimensional lens integral evaluated along the steepest descent manifold $\mathcal{J}_j$ of the classical ray $\bm{x}_j$ assumes the form 
    \begin{align}
        \Psi^{(j)}(\bm{y}) = \left(\frac{\omega}{2\pi i}\right)^{d/2}e^{i \omega T(\bm{x}_j,\bm{y})}\int_{0}^\infty e^{-\omega t}\Delta \phi_{\bm{y}}^{(j)}(t)\mathrm{d}t.\label{eq:Laplace_nD}
    \end{align}
    For more details on the transformation to Laplace-type integrals for multidimensional integrals, we refer to \cite{Howls:1997}.

    \subsection{Transseries}\label{sec:transseries}
    In this section, we proceed beyond the eikonal approximation by extending the seminal work of Dingle \cite{Dingle:1973} to multidimensional integrals. Moreover, we develop an algorithm to efficiently compute the resulting transseries. We first discuss the refractive expansion at nondegenerate classical rays, away from a caustic, before showing how this result extends to caustics using uniform asymptotics.

    \subsubsection{Non-degenerate ray}\label{sec:nondegenerate}
    Starting from the Picard-Lefschetz formula \eqref{eq:PL}, consider the Kirchhoff-Fresnel integral along the steepest descent manifold $\mathcal{J}_j$ associated with the nondegenerate saddle point $\bm{x}_j$,
    \begin{align}
    \Psi^{(j)}(\bm{y}) = \left(\frac{\omega}{2 \pi i}\right)^{d/2} \int_{\mathcal{J}_j} e^{i \omega T(\bm{x}, \bm{y})}\mathrm{d}\bm{x}\,.
    \end{align}
    The Taylor series of the time delay at $\bm{x}_j$ is given by
    \begin{align}
        T(\bm{x},\bm{y}) = T(\bm{x}_j,\bm{y}) + \frac{1}{2}(\bm{x}-\bm{x}_j)^T [\mathcal{H}T(\bm{x}_j,\bm{y})] (\bm{x}-\bm{x}_j) +  \sum_{3 \leq |\alpha|} \frac{(\bm{x}-\bm{x}_j)^\alpha}{\alpha!} \partial^\alpha \varphi(\bm{x}_j)\,,
        \label{eq:taylorMulti}
    \end{align}
    where we use multi-index notation\footnote{
        Given the multi-index $\alpha= (\alpha_1,\dots,\alpha_d)\in \mathbb{N}^d_0$, we define its norm $|\alpha|=\alpha_1+\cdots + \alpha_d$, the first $\alpha!=\alpha_1! \cdots \alpha_d!$ and second factorial $\alpha!!=\alpha_1!! \cdots \alpha_d!!$, the power $\bm{x}^\alpha = x_1^{\alpha_1}\cdots x_d^{\alpha_d}$ of a vector $\bm{x} = (x_1,\dots, x_d)$, and the partial derivative $\partial^\alpha = \partial_{1}^{\alpha_1}\cdots \partial_{d}^{\alpha_d}$.} $\alpha  \in \mathbb{N}^d_0$.
    Expanding the exponential of the third and higher order contributions, we obtain the expansion 
    \begin{align}
            \Psi^{(j)}(\bm{y}) = \left(\frac{\omega}{2 \pi i}\right)^{d/2} e^{i \omega T(\bm{x}_j,\bm{y})}
            \int_{\mathcal{J}_j} e^{\frac{i \omega}{2} \Delta \bm{x}^T [\mathcal{H} T(\bm{x}_j, \bm{y})] \Delta \bm{x}}\sum_{r=0}^\infty \frac{(i \omega)^r}{r!}\left[\sum_{3 \leq |\alpha| } \frac{\Delta \bm{x}^\alpha}{\alpha!} \partial^\alpha \varphi(\bm{x}_j)\right]^r\mathrm{d}\bm{x}\,,
    \end{align}
    with $\Delta \bm{x} = \bm{x} - \bm{x}_j$. Diagonalizing the Hessian $\mathcal{H} T = V \Lambda V^{-1}$ with the eigenvalue matrix $\Lambda = \text{diag}(\lambda_1, \dots, \lambda_d)$ and the eigenvector matrix $V=(\bm{v}_1,\dots,\bm{v}_d)$, and rotating to the eigenframe of the Hessian by setting $\Delta \bm{x} = V \bm{v}$ (leading to the identity $\Delta \bm{x}^T [\mathcal{H} T] \Delta \bm{x} = \bm{v}^T \Lambda \bm{v}$), we use the multinomial expansion to write the power of the sum as a multivariate polynomial 
    \begin{align}
        \left[\sum_{3 \leq |\alpha| } \frac{\Delta \bm{x}^\alpha}{\alpha!} \partial^\alpha \varphi(\bm{x}_j)\right]^r = 
        \left[\sum_{3 \leq |\alpha| } \frac{(V\bm{v})^\alpha}{\alpha!} \partial^\alpha \varphi(\bm{x}_j)\right]^r
        = \sum_{3r \leq |\alpha|} A^r_\alpha \bm{v}^\alpha\,.
        \label{eq:multinomialExpansion}
    \end{align}
    The coefficients $A_\alpha^r$ are determined by the multinomial theorem\footnote{The multinomial theorem expresses the power of a sum as a sum of powers,
        \begin{align}
            (x_1 + x_2  + \cdots + x_m)^n
            = \sum_{\begin{array}{c} k_1+k_2+\cdots+k_m=n \\ k_1, k_2, \cdots, k_m \geq 0\end{array}} {n \choose k_1, k_2, \ldots, k_m}
            x_1^{k_1} \cdot x_2^{k_2} \cdots x_m^{k_m}\,.
        \end{align}
        }
    and capture how the derivatives of the phase variation enter the calculations that follow. Despite raising an infinite sum to a power, each order of the resulting polynomial consists of a finite set of terms. As we will see below, it is of crucial importance that the sum starts at polynomial order $|\alpha|=3r$. Substituting this expansion into the lens integral, we find
    \begin{align}
            \Psi^{(j)}(\bm{y}) 
            &\sim \left(\frac{\omega}{2 \pi i}\right)^{d/2} e^{i \omega T(\bm{x}_j,\bm{y})}
            \sum_{r=0}^\infty \frac{(i \omega)^r}{r!}\sum_{3r \leq |\alpha|}A^r_\alpha
            \int e^{\frac{i \omega}{2} \bm{v}^T \Lambda \bm{v}}  \bm{v}^\alpha \mathrm{d}\bm{v}\label{eq:1}\\
            &=  \frac{e^{i \omega T(\bm{x}_j,\bm{y})}}{\sqrt{\det \nabla \bm{\xi}(\bm{x}_j)}}
            \sum_{r=0}^\infty \frac{1}{r!}
            \sum_{\frac{3r}{2} \leq | \beta|}
            i^{ r + |\beta|}
            \frac{ (2 \beta -1)!!}{ \bm{\lambda}^\beta} 
            A^r_{2\beta} 
                \omega^{r-|\beta|}\,,
    \end{align}
    using the Gaussian identity

    \begin{align}
        \int e^{\frac{i \omega}{2} \bm{v}^T \Lambda \bm{v}}  \bm{v}^\alpha \mathrm{d}\bm{v} = 
        \begin{cases}
            \left(\frac{2 \pi i}{ \omega}\right)^{d/2} \frac{1}{\sqrt{\det \Lambda}}
                \frac{ (\alpha -1)!!}{ \bm{\lambda}^{\alpha / 2}} 
                \left(-i \omega \right)^{ -|\alpha|/2} & \text{when $\alpha \in 2 \mathbb{N}_0^d$,}\\
        0 & \text{otherwise,}
            \end{cases}
    \end{align}
    with the eigenvalue vector $\bm{\lambda}=(\lambda_1, \dots, \lambda_d)$.
    
    Upon ordering terms by powers of $\omega$, we obtain the expansion
    \begin{align}
        \Psi^{(j)}(\bm{y}) &\sim
        \frac{e^{i \omega T(\bm{x}_j,\bm{y})}}{ \sqrt{\det \nabla \bm{\xi}(\bm{x}_j)}} \sum_{m=0}^\infty \frac{T_m^{(j)}(\bm{y})}{\omega^m}\,,
    \end{align}
    with the coefficients given by
    \begin{align}
        T_m^{(j)}(\bm{y}) =
        \sum_{r=0}^{2m} 
        \frac{1}{r!}
        \sum_{|\beta|=m+r}
        i^{r+|\beta|}
        \frac{(2{\beta}-1)!!}{\bm{\lambda}^\beta}
        A^r_{2\beta} \,.\label{eq:nondegenerateExpansion}
    \end{align}
    Each coefficient $T_m^{(n)}$ consists of only a finite set of terms since the sum in \cref{eq:multinomialExpansion} starts at order $|\alpha|=3r$. More specifically, the coefficients $T_0^{(j)}, T_1^{(j)}, \dots, T_M^{(j)}$ for a given order $M$, only depend on the coefficients $A_\alpha^r$ for $r \in \{0,1,\dots, 2M\}$ and $2M \leq |\alpha| \leq 6M$. These coefficients $A_{\alpha}^r$ only depend on the Taylor series \eqref{eq:taylorMulti} to order $|\alpha|=6M$. Using these insights, we may efficiently evaluate the first $M$ coefficients $T_m^{(j)}$ of the asymptotic expansion using \cref{alg:asymptoticTerms}.

    \begin{algorithm}[H]
        \caption{Algorithm for the evaluation of the coefficients $T_m^{(j)}(\bm{y})$ of the asymptotic series.}
        \label{alg:asymptoticTerms}
        \begin{algorithmic}[1]
            \State Evaluate the Taylor series $T(\bm{x},\bm{y})$ at the classical ray $\bm{x}_j$ to order $|\alpha|=6M$
            \begin{align}
                T(\bm{x},\bm{y}) = T(\bm{x}_j,\bm{y}) + \frac{1}{2}(\bm{x}-\bm{x}_j)^T [\mathcal{H}T(\bm{x}_j,\bm{y})] (\bm{x}-\bm{x}_j) +  \sum_{3 \leq |\alpha| \leq 6M} \frac{(\bm{x}-\bm{x}_j)^\alpha}{\alpha!} \partial^\alpha \varphi(\bm{x}_j)\,,
            \end{align}
            with the Hessian $\mathcal{H}T(\bm{x}_j,\bm{y})$, its eigenvalues $\bm{\Lambda}$ and associated eigenvectors $V$.
            \State Raise the truncated Taylor series to the powers $r=0,1,\dots,2M$ and extract coefficients $A_{\alpha}^r$,
            \begin{align}
                \left[\sum_{3 \leq |\alpha| \leq 6M} \frac{(V\bm{v})^\alpha}{\alpha!} \partial^\alpha \varphi(\bm{x}_j)\right]^r = \sum_{3r \leq |\alpha| \leq 6Mr} A^r_\alpha \bm{v}^\alpha\,.
            \end{align}
            For an efficient implementation, we start with $r=0$ and identify $A_\alpha^0=1$ when $\alpha = 0$ and $A_\alpha^1=0$ when $\alpha \neq 0$. Multiply $f=1$ with the sum, $f(\bm{v}) \mapsto f(\bm{v}) \sum_{3 \leq |\alpha| \leq 6M} \frac{(V\bm{v})^\alpha}{\alpha!} \partial^\alpha \varphi(\bm{x}_j)$ and read off the coefficients $A_\alpha^1$. Repeating this process incrementally builds up the coefficients $A_{\alpha}^r$ for $r=0,1,\dots,2M$. 
            \State Evaluate the terms
            \begin{align}
                T_m^{(j)}(\bm{y}) = \sum_{r=0}^{2m} \frac{1}{r!} \sum_{|\beta|=m+r} i^{r+|\beta|} \frac{(2{\beta}-1)!!}{\bm{\lambda}^\beta} A^r_{2\beta}\,,
            \end{align}
            for $m=0,1,\dots,M$ using the coefficients $A_\alpha^r$ obtained in step 2.
        \end{algorithmic}
    \end{algorithm}
    
    To the best of our knowledge, this is the first closed-form expression and algorithm linking the derivatives of the exponent at a nondegenerate saddle point to the associated series for multidimensional Laplace-type integrals. The formula extends the work by Dingle \cite{Dingle:1973}, who focused on the one-dimensional Laplace-type integral and computed the first term for two-dimensional integrals. Moreover, \cref{alg:asymptoticTerms} extends the more recent analysis of one-dimensional Laplace-type integrals by \cite{Nemes:2012}. In \cref{tab:series}, we list the first few terms of the series expansion of the one-dimensional lens. The explicit expression of $T_{m}^{(j)}$ in terms of the derivatives of the time delay grows rapidly. Yet, the algorithm provides a straightforward method for the numerical evaluation of the coefficients. For late terms, when the saddle point $\bm{x}_j$ is only known numerically, the evaluation requires arbitrary-precision floating-point arithmetic to converge. We find this method to be more efficient in practice than the contour integral methods proposed in \cite{Dingle:1973,Berry:1991,Howls:1997}.

    \begin{table}
      \tiny 
        \begin{tabular}{ |c|c| } 
            \hline
            $m$ & $T_m^{(j)}$ \\ 
            \hline
            \rule[-7.5pt]{0pt}{15pt}$0$ & 1  \\ 
            \rule[-7.5pt]{0pt}{15pt}$1$ & $\frac{5 F_3^2-3 F_2 F_4}{24 F_2^3}$  \\ 
            \rule[-7.5pt]{0pt}{15pt}$2$ & $\frac{385 F_3^4-630 F_2 F_4 F_3^2+168 F_2^2 F_5 F_3+3 F_2^2 \left(35 F_4^2-8 F_2 F_6\right)}{1152 F_2^6}$  \\ 
            \rule[-7.5pt]{0pt}{15pt}$3$ & $\frac{425425 F_3^6-1126125 F_2 F_4 F_3^4+360360 F_2^2 F_5 F_3^3-10395 F_2^2 \left(8 F_2 F_6-65 F_4^2\right) F_3^2+3240 F_2^3 \left(4 F_2 F_7-77 F_4 F_5\right) F_3-27 F_2^3 \left(1925 F_4^3-840 F_2 F_6 F_4+8 F_2 \left(5 F_2 F_8-63 F_5^2\right)\right)}{414720 F_2^9}$  \\ 
            \hline
        \end{tabular}
        \normalsize
        \caption{The first few terms of the asymptotic series $T_m^{(j)}$ of the integral $\int e^{-\omega f(\bm{x})}\mathrm{d}\bm{x}$ with $F_i = \partial_x^n f(x_j)$ at the saddle point $x_j$ of $f$.}\label{tab:series}
    \end{table}

    Substituting the expansion for $\Psi^{(j)}(\bm{y})$ in the Picard-Lefschetz formula yields the transseries for the lens integral
    \begin{align}
        \Psi(\bm{y}) &\sim \sum_j n_j(\bm{y}) \Psi^{(j)}(\bm{y}) \\
        &=\sum_{j}n_j(\bm{y}) \frac{e^{i \omega T(\bm{x}_j,\bm{y})}}{ \sqrt{\det \nabla \bm{\xi}(\bm{x}_j)}} \sum_{m=0}^\infty \frac{T_m^{(j)}(\bm{y})}{\omega^m}\,.
        \label{eq:transseries}
    \end{align}
    
\begin{figure}
        \centering
        \includegraphics[width=0.4\textwidth]{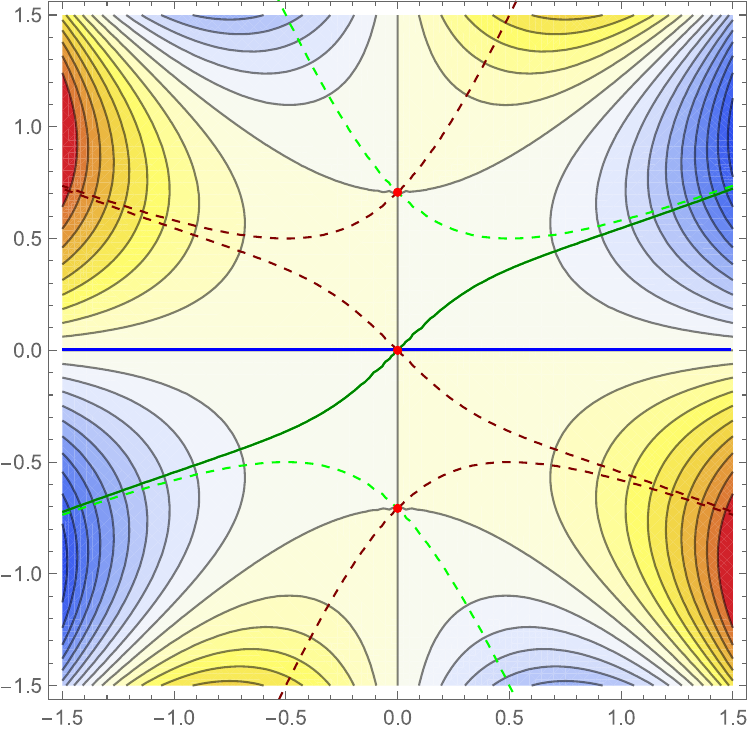}
        \caption{The Picard-Lefschetz deformation of the original integration domain of the integral $\int_\mathbb{R} e^{i \omega(x^2+x^4)}\mathrm{d}x$ in the complex $x$ plane. The analytic continuation of the exponent has three saddle points (the red points) with the associated steepest descent and ascent manifolds (the blue and red curves). The optimal deformation (the light green curve) of the original integration domain $\mathbb{R}$ (the dark green line).}\label{fig:appendix_PL}
    \end{figure}

    The transseries is a formal object that bundles the saddle points, through exponential terms, together with their associated asymptotic series. Transseries not only emerge in the study of integrals but also arise in the perturbative study of differential equations (see, for example, \cite{Daalhuis:1998b}). We calculate the transseries for the quartic integral:
    \begin{example}\label{ex:quartic_trans}
        The quartic integral 
        \begin{align}
            \Phi = \sqrt{\frac{\omega}{\pi i}}\int_{-\infty}^\infty e^{i \omega (x^2 + x^4)}\mathrm{d}x,
        \end{align}
        has three saddle points located along the imaginary axes
        \begin{align}
            x = 0\,, \quad x = \pm i / \sqrt{2}\,.
        \end{align}
        Tracing the steepest descent contours, we find that only the real saddle point is relevant to the integral (see fig.\ \ref{fig:appendix_PL}). The series expansion around the saddle $x=0$ coincides with the diffractive expansion \cref{eq:quartic_expansion} in \cref{ex:quartic} and the series expansion at the saddle points $\pm i / \sqrt{2}$ is given by
        \begin{align}
            \sqrt{\frac{\omega}{\pi i}} \int e^{\omega(-i/4-2 i x^2 \mp 2 \sqrt{2} x^3 + i x^4)} \mathrm{d}x&=\sqrt{\frac{\omega}{\pi i}}e^{-\frac{i \omega}{4}}
            \sum_{n=0}^\infty \frac{\omega^n}{n!} \int_{-\infty}^\infty e^{-2 i \omega x^2} \left[ \mp 2\sqrt{2} x^3 + i x^4\right]^n\mathrm{d}x \nonumber\\
            &= \frac{e^{-\frac{i \omega}{4}} }{\sqrt{2}i}
            \left[1-\frac{3i}{16 \omega^2} + \frac{15 i}{8 \omega^3} -\frac{105}{256 \omega^4} + \dots \right]\,.
        \end{align}
        This leads to the transseries
        \begin{align}
            \Phi \sim n_{x=0} \sum_{n=0}^\infty \frac{(4 n - 1)!!}{4^{ n} n! i^n \omega^n}  + \left(n_{x=- i / \sqrt{2}} + n_{x= i / \sqrt{2}}\right) \frac{e^{-\frac{i \omega}{4}} }{\sqrt{2} i } \left[1-\frac{3i}{16 \omega^2} + \frac{15 i}{8 \omega^3} -\frac{105}{256 \omega^4} + \dots \right]\,,
        \end{align}
        with the intersection numbers $n_{x=0}=1$ and $n_{x=\pm i/\sqrt{2}}=0$ when starting with the integral along the real line.
    \end{example}
    
    To zeroth order, the transseries reduces to the eikonal approximation as expected. Contrary to what one may hope, the sum $\sum_{m=0^\infty} T_m^{(j)}/\omega^m$ diverges for each classical ray (as we already observed in \cref{ex:quartic}) since exchanging the infinite sum with the integration symbol in \cref{eq:1} creates an asymptotic rather than a convergent series. In \cref{sec:resurgence}, we outline a rigorous interpretation of these divergent transseries and discuss ways to approximate the lens integral to an arbitrary level of accuracy.

    \subsubsection{Caustics rays}\label{sec:uniform}
    At a caustic, both the eikonal and resurgence analysis (to be discussed later in \cref{sec:resurgence}) of the transseries break down since the determinant of the deformation tensor vanishes. The time delay function in the immediate vicinity of the caustic is no longer described by the Hessian, and the derivation in \cref{sec:nondegenerate} does not apply. Fortunately, caustics are classified in terms of the elementary catastrophes and can be treated separately. The seven elementary catastrophes form a complete classification of degenerate critical points up to four external parameters. In \cref{tab:catastrophes} we list the first seven elementary catastrophes and their unfolding, describing their canonical behaviour. For a review on catastrophe theory more generally, see \cite{Saunders:1980,Arnold:1985}. We will derive the uniform asymptotic expansion of the fold caustic here. The higher-order caustics follow analogously. For more details on uniform asymptotic asymptotics we refer to \cite{Chester:1957, Berry:1989, Olver:1991, Berry:1993, Berry:1994, Daalhuis:1994, Daalhuis:2000, Khwaja:2013, Temme:2013, Khwaja:2016}. For a first-order application of uniform asymptotics to wave optics in astronomy, see 
    \cite{Grillo:2018}.

    \begin{table}
        \begin{tabular}{|c|c|c|l|} 
            \hline
            \rule[-5pt]{0pt}{15pt} Name                 & Symbol  & Dimensionality      & Unfolding\\
            \hline
            \rule[-5pt]{0pt}{15pt} Fold                 & $A_2$   & 1D                  & $\psi(t) = \frac{1}{3} t^3 + \mu t$  \\ 
            \rule[-7.5pt]{0pt}{15pt} Cusp               & $A_3$   & 2D                  & $\psi(t) = \frac{1}{4}t^4 + \frac{\mu_2}{2} t^2 + \mu_1 t$  \\ 
            \rule[-7.5pt]{0pt}{15pt} Swallowtail        & $A_4$   & 3D                  & $\psi(t) = \frac{1}{5} t^5 + \frac{\mu_3}{3} t^3 + \frac{\mu_2}{2}t^2 + \mu_1 t$  \\ 
            \rule[-7.5pt]{0pt}{15pt} Butterfly          & $A_5$   & 4D                  & $\psi(t) = \frac{1}{6} t^6 + \frac{\mu_4}{4}t^4 + \frac{\mu_3}{3} t^3 + \frac{\mu_2}{2}t^2 + \mu_1 t$   \\ 
            \rule[-7.5pt]{0pt}{15pt} Elliptic umbilic   & $D_4^-$ & 3D                  & $\psi(\bm{t}) = t_1^3 - 3 t_1 t_2^2 - \mu_3(t_1^2+t_2^2) - \mu_2 t_2 - \mu_1 t_1$  \\ 
            \rule[-7.5pt]{0pt}{15pt} Hyperbolic umbilic & $D_4^+$ & 3D                  & $\psi(\bm{t}) = t_1^3 + t_2^3 - \mu_3 t_1 t_2 - \mu_2 t_2 - \mu_1 x_1$ \\ 
            \rule[-7.5pt]{0pt}{15pt} Parabolic umbilic  & $D_5$   & 4D                  & $\psi(\bm{t}) = t_1^4 + t_1 t_2^2 + \mu_4 t_2^2 + \mu_3 x_1^2 + \mu_2 t_2 + \mu_1 t_1$\\ 
            \hline
        \end{tabular}
        \caption{The unfoldings $\psi$ of the elementary catastrophes, with the intrinsic variables labelled by $\bm{t}$ and the external parameters labelled by $\bm{\mu}$.}\label{tab:catastrophes}
    \end{table}
    
    For convenience, we will first consider the one-dimensional lens problem. In the following, we will only consider the transseries contributions of the rays involved in the caustic. When $x_c$ is a fold catastrophe, mapped to the fold caustic at $y_c$ in the image plane, both its first- and second-order derivatives vanish, \textit{i.e.}, the point $(x_c,y_c)$ is a real solution of the equations
    \begin{align}
        \partial_x T(x_c,y_c) &= x_c - y_c + \varphi'(x_c) = 0\,,\\
        \partial_x^2 T(x_c, y_c) &= \frac{1+\varphi''(x_c)}{2} = 0\,.
    \end{align}
    The Taylor expansion of the time delay function at the fold caustic then assumes the form 
    \begin{align}
        T(x,y) = 
        T(x_c,y)
        - \Delta y \Delta x
        + \sum_{n=3}^\infty \frac{\varphi^{(n)}(x_c)}{n!}\Delta x^n\,,
    \end{align}
    with $x - x_c = \Delta x$ and $y-y_c = \Delta y$. The lens integral,
    \begin{align}
        \Psi(y) = \sqrt{\frac{\omega}{2 \pi i}} e^{i \omega T(x_c, y)} \int e^{i \omega\left(- \Delta y \Delta x
        + \sum_{n=3}^\infty \frac{\varphi^{(n)}(x_c)}{n!}\Delta x^n\right)}\mathrm{d} \Delta x\,,
    \end{align}
    simplifies using the coordinate transformation $\Delta x = \sqrt[3]{\frac{2}{\varphi^{(3)}(x_c)}}t$ to an expansion in terms of the canonical form of the fold catastrophe, 
    \begin{align}
        \Psi(y) &\sim \sqrt{\frac{\omega}{2 \pi i}} \sqrt[3]{\frac{2}{\varphi^{(3)}(x_c)}} e^{i \omega T(x_c, y)} \int e^{i \omega \left(\frac{1}{3} t^3 + \mu t + \sum_{n=4}^\infty \frac{B_n}{n!} t^n\right)}\mathrm{d}t\\
        &=
         \sqrt{\frac{\omega}{2 \pi i}} \sqrt[3]{\frac{2}{\varphi^{(3)}(x_c)}} e^{i \omega T(x_c,y )}\int e^{i\omega \left(\frac{1}{3}t^3+ \mu t\right)} 
         \sum_{r=0}^\infty \frac{(i\omega)^r}{r!}
        \left[ \sum_{n=4}^\infty \frac{B_n}{n!} t^n\right]^r \mathrm{d}t\,,
    \end{align}
    with 
    \begin{align}
        \mu =  -\sqrt[3]{\frac{2}{\varphi^{(3)}(x_c)}} \Delta y \,, \quad
        B_n=\varphi^{(n)}(x_c) \left(\frac{2}{\varphi^{(3)}(x_c)}\right)^{n/3}\,.
    \end{align}
    Expanding the power of the sum, 
    \begin{align}
        \left[ \sum_{n=4}^\infty \frac{B_n}{n!} t^n\right]^r = \sum_{p=4r}^\infty A_p^r t^p\,,
    \end{align}
    we obtain the series 
    \begin{align}
        \Psi(y) \sim \sqrt{\frac{\omega}{2 \pi i}} \sqrt[3]{\frac{2}{\varphi^{(3)}(x_c)}} e^{i \omega T(x_c,y )}
         \sum_{r=0}^\infty \frac{(i\omega)^r}{r!}
        \sum_{p=4r}^\infty A_p^r \int e^{i\omega \left(\frac{1}{3}t^3+ \mu t\right)} t^p \mathrm{d}t\,.
    \end{align}
    This expansion can then be expressed in terms of Airy functions as
    \begin{align}
        \int e^{i \omega\left(\frac{1}{3}t^3 + \mu t\right)} t^p \mathrm{d}t 
        =
        \frac{2 \pi }{i^p \omega^{(p+1)/3}}\Ai^{(p)}(\omega^{2/3} \mu)\,,
    \end{align}
    where $\Ai^{(p)}(x)$ denotes the $p$th-order derivative of the Airy function. Working in terms of the external parameter $\tilde{\mu} = \omega^{2/3}\mu$, 

    the uniform asymptotic expansion assumes the form 
    \begin{align}
        \Psi(y) 
        \sim 
             \sqrt{\frac{\omega}{2 \pi i}} \sqrt[3]{\frac{2}{\varphi^{(3)}(x_c)}} e^{i \omega T(x_c,y )}
        \sum_{m=0}^\infty
        \frac{T_m}{\omega^{m/3}}\,,
    \end{align}
    with 
    \begin{align}
        T_m = \sum_{r=0}^{m-1} \frac{2 \pi i^{-2r-m+1} A_{3r+m-1}^r }{r!} 
            \Ai^{(3r +m - 1)}( \tilde{\mu})\,.
    \end{align}
    
    Using the identity $\Ai''(x) = x \Ai(x)$, we can express the $p$th-order derivative in terms of the Airy function and its first-order derivative (for a closed-form expansion see \cite{Abramochkin:2018}). This leads to a formulation of the uniform expansion of the fold caustic in terms of the Airy function and its first-order derivative,
   \begin{align}
        \Psi(y) \sim \sqrt{-2 \pi i \omega} \sqrt[3]{\frac{2}{\varphi^{(3)}(x_c)}}  e^{i \omega T(x_c, y)}  \left[
            \Ai(\tilde{\mu})
        \sum_{n=1}^\infty \frac{a_n}{\omega^{n/3}}  
        +   \Ai'(\tilde{\mu}) 
        \sum_{n=1}^\infty \frac{b_n}{\omega^{n/3}} \right]\,.
    \end{align}
    In \cref{tab:foldasymptotics} we present explicit expressions for the first few terms in the asymptotic expansion. As for the transseries obtained in \cref{sec:nondegenerate}, the asymptotic series of the uniform asymptotic expansion first converges before eventually diverging. Near higher-order caustics, an analogous uniform asymptotic expansion follows in terms of a set of canonical diffraction integrals $\int e^{i \psi(\bm{t})}\mathrm{d}\bm{t}$
    with the associated unfolding $\psi(\bm{t})$ (see \cref{tab:catastrophes}) of the catastrophe in question \cite{Berry:1976}. For the cusp caustic, this integral is related to the famous Pearcey integral \cite{Pearcey:1946}. Practically, the efficient evaluation of the coefficients follows \cref{alg:asymptoticTerms}.

\begin{figure}
        \centering
        \begin{subfigure}[b]{0.32\textwidth}
            \includegraphics[width=\textwidth]{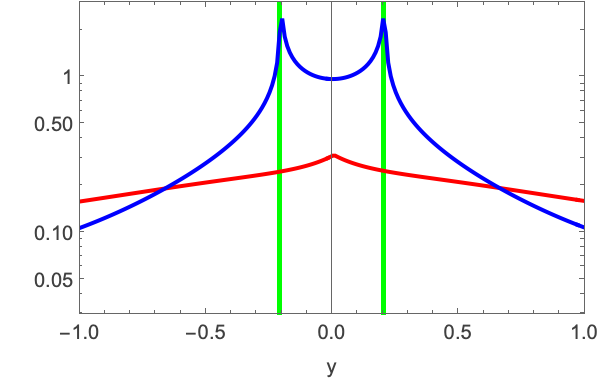}
            \caption{$\omega=1$}
        \end{subfigure}
        \begin{subfigure}[b]{0.32\textwidth}
            \includegraphics[width=\textwidth]{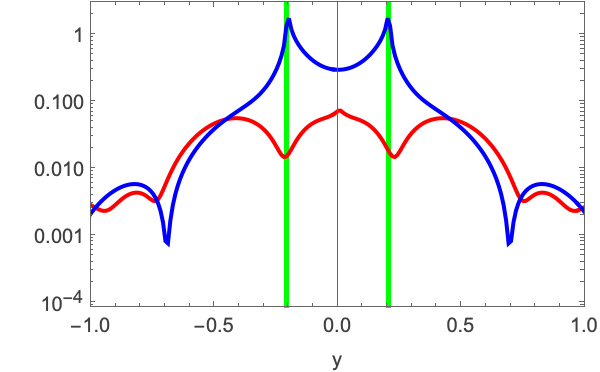}
            \caption{$\omega=10$}
        \end{subfigure}
        \begin{subfigure}[b]{0.32\textwidth}
            \includegraphics[width=\textwidth]{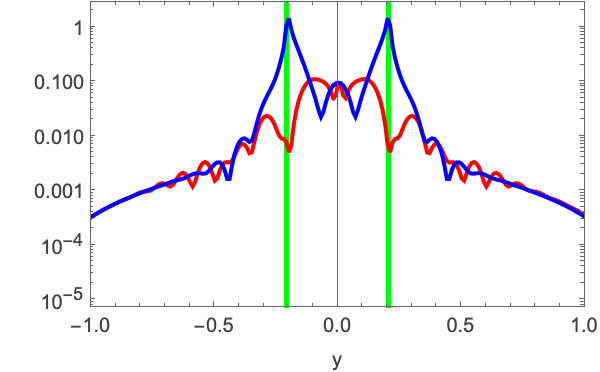}
            \caption{$\omega=50$}
        \end{subfigure}
        \caption{The difference between the exact lens integral and the leading uniform approximation $|\Psi(y) - \Psi_{\text{uniform}}(y)|$ (red) and the eikonal approximation $|\Psi(y) - \Psi_{\text{eikonal}}(y)|$(blue)  for the one-dimensional Lorentzian lens with amplitude $\alpha =1$. The green lines mark the caustics.}\label{fig:uniform}
    \end{figure}

\begin{figure}
        \centering
        \begin{subfigure}[b]{0.49\textwidth}
            \includegraphics[width=\textwidth]{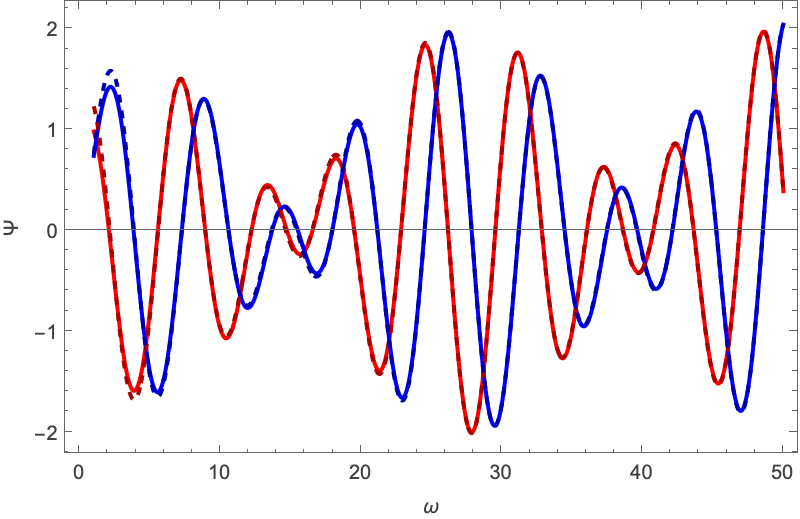}
            \caption{The lens amplitude}
        \end{subfigure}
        \begin{subfigure}[b]{0.49\textwidth}
            \includegraphics[width=\textwidth]{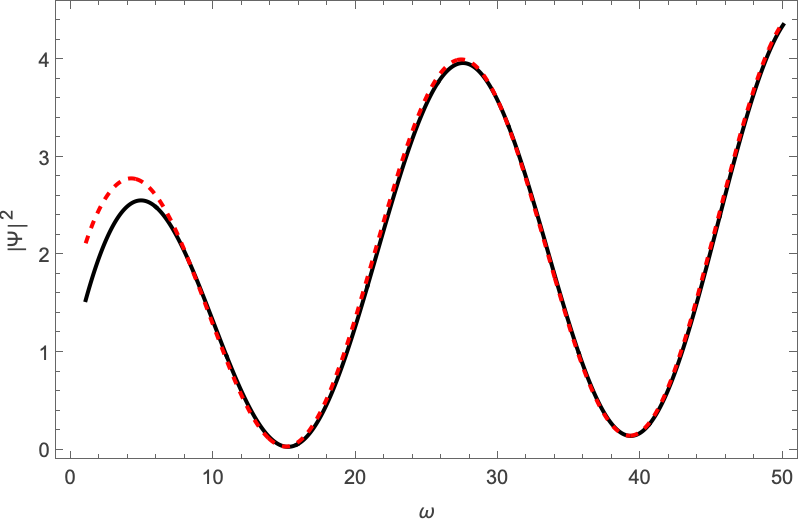}
            \caption{The intensity}
        \end{subfigure}
        \caption{The lens amplitude and intensity of the one-dimenisonal Lorentzian lens with $\alpha =1$ at the fold caustic and the uniform approximation as a function of frequency. \text{Left:} The real (red) and imaginary parts (blue) of the lens amplitude. The solid curves show the true amplitude and the dashed curves show the leading uniform approximation. \text{Right:} The intensity of the lens (black) and the leading uniform approximation (dashed red).}\label{fig:uniform_omega}
    \end{figure}

    To leading order, the uniform approximation is given by
       \begin{align}
        \Psi_{\text{uniform}}(y) =  
        \sqrt{-2 \pi i}\sqrt[3]{\frac{2}{\varphi^{(3)}(x_c)}}  e^{i \omega T(x_c, y)}  
         \omega^{1/6} \Ai(\omega^{2/3}\mu)\,.
    \end{align}
    See \cref{fig:uniform,fig:uniform_omega} for a numerical demonstration of the zeroth-order uniform approximation of the one-dimensional Lorentzian lens model. Away from the caustics, the standard eikonal approximation provides a more accurate description than the uniform approximation. In the vicinity of the caustics, however, the uniform approximation smoothly approaches the lensing amplitude and regularises the divergence inherent in the eikonal approximation. As we increase the frequency, the region around the caustics for which the uniform approximation is beneficial shrinks. The uniform approximation works best in the refractive regime. In the diffractive regime, the uniform approximation starts to deviate from the true lens amplitude.
    To improve the convergence properties of the transseries \eqref{eq:transseries} near the caustics, we replace the terms associated with the caustic with the associated uniform asymptotic formula.  

    The application of the uniform approximation near the fold caustics in astrophysical lensing was recently introduced by \cite{Grillo:2018}. However, their treatment differs significantly from the present discussion in several respects. Firstly, they focus on the leading contribution and do not consider the full asymptotic series. Secondly, using the form $\Psi = \sqrt{2 \pi} e^{i \chi} [g_1 \Ai(\zeta) + g_2 \Ai'(\eta)]$ they approximate the functions $g_1, g_2, \chi$ and $\zeta$ from the geometric optics approximation rather than from an expansion of the time delay function at the fold caustic. Consequently, their treatment neatly extends the eikonal approximation to the fold caustics, but does not relate it directly to the Taylor series of the time delay at the fold caustic.

    \begin{table}
        \begin{tabular}{|c|c|c|} 
            \hline
             $n$              & $a_n$  & $b_n$\\
             \hline
             \rule[-5pt]{0pt}{15pt}$1$ & $1$ &  $0$\\
             \rule[-7.5pt]{0pt}{15pt}$2$ & $\frac{i B_4 \tilde{\mu} ^2}{24} $ & $\frac{i B_4}{12}$\\
             \rule[-7.5pt]{0pt}{15pt}$3$ & $\frac{B_5 \tilde{\mu} }{30}-\frac{B_4^2 \tilde{\mu}  \left(\tilde{\mu} ^3+28\right)}{1152}$ & $\frac{\left(4 B_5-5 B_4^2\right) \tilde{\mu} ^2}{480} $\\  
             \rule[-7.5pt]{0pt}{15pt}$4$ & $-\frac{i \left(-576 B_5 B_4 \left(4 \tilde{\mu} ^3+7\right)+576 B_6 \left(\tilde{\mu} ^3+4\right)+5 B_4^3 \left(\tilde{\mu} ^6+260 \tilde{\mu} ^3+280\right)\right)}{414720}$ & $-\frac{i \tilde{\mu}  \left(25 B_4^3 \left(\tilde{\mu} ^3+20\right)-24 B_5 B_4 \left(\tilde{\mu} ^3+52\right)+576 B_6\right)}{69120}$\\
            \hline
        \end{tabular} 
        \caption{The first terms in the asymptotic series of the uniform asymptotic expansion of a fold caustic.}\label{tab:foldasymptotics}
    \end{table}

    A typical caustic is crossed in the overfocussed regime, corresponding
    to large $\alpha$.  The asymptotic peak intensity of this image is $I=I_0
    \omega^{1/3}$ with $I_0=\frac{ 2^{1/3} 16\pi}{27 (3\alpha)^{2/3}
      \Gamma[2/3]^2}$ and $I \sim 0.61 (\omega/\alpha^2)^ {1/3}$.  For off
    axis lenses, the peak caustic flux is small.  In typical ESE scenarios
    (where $\alpha$ is called convergence $\kappa$ in \cite{Zhu}), $\kappa \sim 20$, $\omega\sim 40$, the peak caustic flux is
    only 30\% of the unlensed flux, consistent with the lack of salient
    caustic peaks in secondary wavefields.  The average geometric optics 
    lensed image flux is $\propto 1/\kappa$, leading to a modeset fractional 
    caustic enhancement $\propto (\omega\kappa)^{1/3}$.

   \bigskip

    For multidimensional integrals, the asymptotic series follows from a combination of the derivation of the degenerate and non-degenerate expansions. For example, for a fold caustic in a two-dimensional lens integral, for which one of the eigenvalues of the deformation tensor vanishes, there exists a coordinate system in which the time delay takes the form
    \begin{align}
        T(\bm{x},\bm{y}) \approx T(\bm{x}_c, \bm{y}) + \frac{1}{3} \Delta x_1^3 + \mu(\bm{y}) \Delta x_1 + \Delta x_2^2 
    \end{align}
    to leading order. Upon expanding the exponential of the higher-order terms, the integrals in the expansion factorise. The integral over $x_1$ follows the derivation of the fold caustic, while the integral over $x_2$ follows the derivation of the non-degenerate saddle point.

    \section{Resurgence}\label{sec:resurgence}
    In \cref{sec:transseries}, we extended the eikonal approximation to the transseries expansion. At first glance, this appears to offer little improvement, since each series associated with a classical ray is asymptotic and divergent! In contrast to the diffractive expansion, the formal transseries does therefore naively fail to approximate the lens integral. This puzzling behaviour historically confounded several eminent mathematicians -- among them Niels Abel, who famously remarked that “divergent series are the invention of the devil, and it is shameful to base on them any demonstration whatsoever.” Bayes claimed that Stirling’s series, which is indeed asymptotic, “can never properly express any quantity at all” and the methods used to obtain it “are not to be depended upon.” Yet, over the past century, the pioneering work of Borel, Dingle, Écalle, and later Berry, Howls, and Olde Daalhuis have provided a rigorous framework for the analysis of asymptotic series, known as the theory of resurgence \cite{Dingle:1973, Ecalle:1981, Berry:1990, Ecalle:1993, Berry:1991,Daalhuis:1996, Howls:1997, Daalhuis:1998, Daalhuis:1999}. Their contributions reveal how transseries can be resummed, yielding exponentially accurate approximations to the underlying integrals or, more generally, solutions to differential equations. Interestingly, while the diffractive expansion first diverges before converging to the lens amplitude, the refractive approximation -- through resurgence on the formal transseries --  can be made to converge more quickly and unveils the analytic structure of the interference pattern. In this section, we discuss the superasymptotic approximation, Borel resummation and ultimately the hyperasymptotic approximation that allows us to extract finite results for the lens amplitude from a formal transseries.

    \subsection{The superasymptotic approximation}\label{sec:superasymptotics}
    The superasymptotic approximation is the most direct and practical way to work with the transseries and it has a rich history going back to work by Poincar\'e, Stieltjes, Cauchy and Stokes. The superasymptotic method was formalised by Berry and Howls \cite{Berry:1990}. As illustrated in \cref{ex:quartic}, an asymptotic series typically approaches the true result before eventually diverging. The reason underlying this behaviour will become clear in \cref{sec:hyperasymptotics}. It is thus natural to truncate the sum $\sum_{m=0}^\infty T_m^{(j)}/\omega^m$ at the index for which $|T_m^{(j)}/\omega^m|$ is the smallest, denoted as $N^{(j)}$. This leads to the superasymptotic approximation 
    \begin{align}
        \Psi_{\text{super}}(\bm{y}) 
        =\sum_{j}n_j(\bm{y}) \frac{e^{i \omega T(\bm{x}_j,\bm{y})}}{ \sqrt{\det \nabla \bm{\xi}(\bm{x}_j)}} \sum_{m=0}^{N^{(j)}} \frac{T_m^{(j)}(\bm{y})}{\omega^m}\,.
    \end{align}
    As we will discuss in \cref{sec:hyperasymptotics}, the optimal truncation point $N^{(j)}$ is proportional to the difference of the time delay in the ray $\bm{x}_j$ and the closest classical ray in the Borel plane. More explicitly, defining the so-called singulant 
    \begin{align}
        F_{jk} = i\left(T(\bm{x}_j,\bm{y})-T(\bm{x}_k,\bm{y})\right)\,.
    \end{align}
    the optimal truncation point $N^{(j)}$ is the integer part of $|\omega F_{jk^*}|$ with $k^*$ the label of the ray $\bm{x}_{k^*}$ that lives on the same Riemann sheet and is closest to $\bm{x}_j$ in the Borel plane, \textit{i.e.} the ray on the same Riemann sheet in the Borel plane for which the modulus of the singulant $|F_{jk}|$ is the smallest. The error in the superasymptotic approximation is of the order $e^{-|\omega F_{jk^*}|}$.

\begin{figure}
        \centering
        \begin{subfigure}[b]{0.45\textwidth}
            \includegraphics[width=\textwidth]{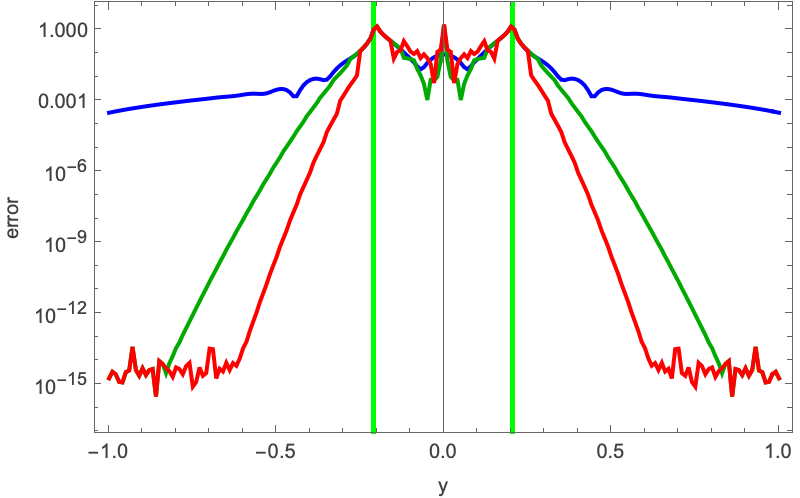}
            \caption{The error as a function of space}
        \end{subfigure}
        \begin{subfigure}[b]{0.45\textwidth}
            \includegraphics[width=\textwidth]{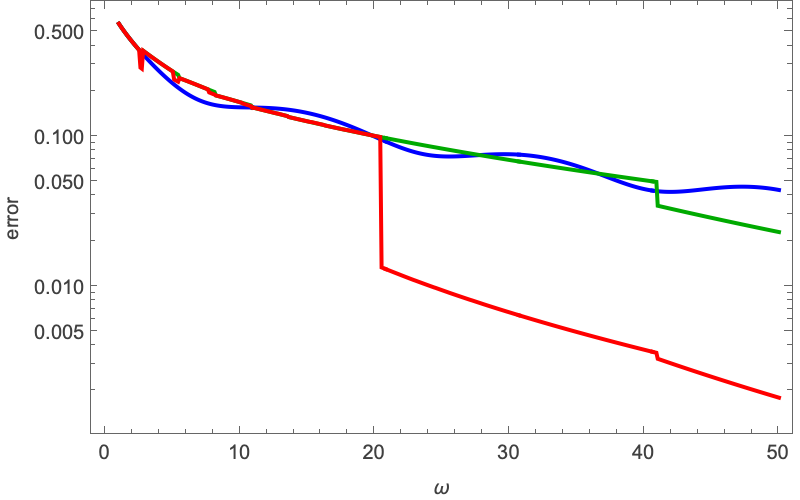}
            \caption{The error as a function of frequency}
        \end{subfigure}
        \caption{\textit{Left:} The modulus of the lens amplitude (black), eikonal approximation (blue), and superasymptotic approximation (green), and first-order hyperasymptotic approximation (black) of the one-dimensional Lorentzian lens as a function of the frequency with the amplitude $\alpha=1$ and the point on the sky $y=0.3$ as a function of the frequency. \textit{Right:} The associated error with respect to the eikonal approximation $|\Psi - \Psi_{\text{eikonal}}|$ (blue), the superasymptotic approximation $|\Psi - \Psi_{\text{super}}|$ (green) and first-order hyperasymptotic approximation $|\Psi - \Psi_{\text{hyper}}|$ (red).}\label{fig:hyper_omega}
    \end{figure}

    The superasymptotic approximation provides a natural refinement of the eikonal approximation. We compare the error of the eikonal and the superasymptotic approximation of the one-dimenisonal Lorentzian lens as a function of space $\bm{y}$ and frequency $\omega$ in \cref{fig:hyper_omega}. Away from the caustics, the superasymptotic approximation provides a significant increase in accuracy. In the one-dimensional Lorentzian lens example, the superasymptotic approximation reaches the accuracy level with which we numerically evaluate the integral with a Picard-Lefschetz integrator for $|y| \geq 0.8$  (see \cref{sec:PL}). Near the caustics, where $|\omega F_{jk^*}|$ drops below unity, the superasymptotic approximation coincides with the eikonal approximation. In these regions the asymptotic series directly diverges. The same phenomena are observed in the frequency domain. For large frequencies, the superasymptotic approximation is a significant improvement over the eikonal approximation. For small frequencies, when $|\omega F_{jk^*}|$ drops below unity, the superasymptotic approximation coincides with the eikonal approximation. Near caustics, the transseries obtained from the expansion around nondegenerate rays fails. As we observed in \cref{sec:uniform}, in these regions the superasymptotic approximation of the uniform asymptotic expansion provides a significantly better approximation.

    \subsection{Borel resummation}\label{sec:borel}
    At first sight, the superasymptotic approximation appears to encounter an impenetrable barrier. We leveraged the convergent part of each asymptotic expansion and are left with a divergent tail! Yet, the divergent tail turns out to encode the full analytic structure of the oscillatory integral (this idea was first realized by Dingle \cite{Dingle:1973}). To understand this in detail, we will first show how the formalism of Borel resummation can, in principle, recover the lens integral \cite{Borel:1899}. In \cref{sec:hyperasymptotics}, we use hyperasymptotics to approximate the lens integral to arbitrary precision beyond the superasymptotic approximation.

    Starting with the transseries, 
    \begin{align}
        \Psi(\bm{y}) \sim \sum_{j}n_j(\bm{y}) \frac{e^{i \omega T(\bm{x}_j,\bm{y})}}{ \sqrt{\det \nabla \bm{\xi}(\bm{x}_j)}} \sum_{m=0}^\infty \frac{T_m^{(j)}(\bm{y})}{\omega^m}\,,\label{eq:trans2}
    \end{align}
    we Borel resum the asymptotic series $T^{(j)}_{\bm{y}}(\omega)=\sum_{m=0}^\infty T_m^{(j)}(\bm{y}) / \omega^m$ for each classical ray with the following scheme:
    
    \begin{enumerate}
        \item We first define the Borel transform by dividing each term by a factorial
        \begin{align}
            \mathcal{B}T^{(j)}_{\bm{y}}(t) = \sum_{m=0}^\infty \frac{T^{(j)}_{m}(\bm{y})}{m!}t^m\,.\label{eq:BorelTransform}
        \end{align}
        The series $\sum_{m=0}^\infty T_m^{(j)}/\omega^m$ is said to be Borel summable when the Borel transform $\mathcal{B}T^{(j)}_{\bm{y}}$ converges. Asymptotic series originating from Laplace-type integrals generally diverge factorially, making the Borel transform converge.
        \item 
        Next, we Laplace transform the Borel transform to obtain the so-called Borel sum of the asymptotic series
        \begin{align}
            T_B^{(j)}(\bm{y}) = \omega \int_0^\infty e^{-\omega t} \mathcal{B}T^{(j)}(t) \mathrm{d}t\label{eq:BorelSummation}\,.
        \end{align}
    \end{enumerate}
    Now, rather surprisingly, replacing the asymptotic series $\sum_{m=0}^\infty T_m^{(j)}/\omega^m$ with the Borel resummed series $T_B^{(j)}$ yields the full lens amplitude 
      \begin{align}
        \Psi(\bm{y}) = \sum_{j}n_j(\bm{y}) \frac{e^{i \omega T(\bm{x}_j,\bm{y})}}{ \sqrt{\det \nabla \bm{\xi}(\bm{x}_j)}} T_B^{(j)}(\bm{y})\,.
    \end{align}
    The Borel resummation method has converged the formal transseries into the lens amplitude, irrespective of the frequency. We illustrate Borel resummation for the asymptotic series of the quartic integral.
    \begin{example}
        In \cref{ex:quartic,ex:quartic_trans}, we derived that the diffractive expansion of the quartic integral leads to the asymptotic series
        \begin{align}
             \Psi = \sqrt{\frac{\omega}{2 \pi i}} \int_{-\infty}^\infty e^{i\omega (x^2 + x^4)}\mathrm{d}x 
            \sim \sum_{n=0}^\infty \frac{(4 n - 1)!!}{4^{ n} n! i^n \omega^n} \,.
        \end{align}
        Let us Borel resum this divergent series and compare it with the closed form expression presented in \cref{ex:quartic}. The series diverges factorially, and the Borel transform converges 
        \begin{align}
            \mathcal{B} \Psi(t) &= \sum_{n=0}^\infty \frac{(4 n - 1)!!}{4^{ n} (n!)^2 i^n} t^n\\
            &= \frac{2 }{\pi  \sqrt[4]{1+4 i t}}K\left(\frac{1}{2}-\frac{1}{2 \sqrt{4 i t+1}}\right)
        \end{align}
        in terms of the complete elliptic integral of the first kind $K(m)$. The Laplace transform recovers the original integral 
        \begin{align}
            \omega \int_0^\infty e^{-\omega t} \mathcal{B}I(t)\mathrm{d}t 
            = 
            \sqrt{\frac{\omega}{4\pi i}} e^{-\frac{i \omega}{8}} K_{1/4}\left(-\frac{i \omega}{8}\right)\,,
        \end{align}
        where $K_n(x)$ denotes the modified Bessel function of the second kind. We have thus evaluated the quartic integral by means of Borel resummation using the closed form representation of the Borel transform.
    \end{example}

    Let us now make some informal observations:
    \begin{itemize}
        \item At first sight, Borel resummation is a remarkable procedure. We artificially introduce a factorial suppression and subsequently `integrated it out'. The central idea being that the Laplace transform of the monomial factor leads to a multiplication by the same factorial factor
        \begin{align}
            \int_0^\infty e^{-t} t^n = n!\,.
        \end{align}
        Indeed, when a series $\sum_{n=0}^\infty a_n$ converges absolutely -- that is to say $\sum_{n=0}^\infty |a_n| < \infty$ -- we can change the order of summation and integration. Moving the Laplace integral inside the sum, the Laplace integral cancels the factorial in the Borel transform. Consequently, Borel resummation on an absolutely convergent series recovers the original series.
        \item On the other hand, when truncating the sum in the Borel transform, the Laplace transform recovers the original series irrespective of whether it converges or diverges. Having complete knowledge of the transseries and performing the infinite sum in the Borel transform before the Laplace transform is critical.
        \item In general -- when the sum does not converge absolutely -- exchanging the integration symbol with the infinite sum is not admissible. The order of operations matters. In a loose sense, the divergence in \cref{eq:trans2} arose from moving the infinite sum outside the integral, ignoring the conditions of the dominated convergence theorem. The original integral is recovered by inserting the factorial and `cancelling' this step with the Laplace transform! The question we now turn to is what properties, in these cases, guarantee convergence to the original integral?
    \end{itemize}

    \bigskip
    To understand the idea underlying Borel resummation, we note that each analytic function $f(\epsilon)$ has a unique expansion 
    \begin{align}
        f(\epsilon) = \sum_{n=0}^\infty f_n \epsilon^n
    \end{align}
    in the small parameter $\epsilon$. The converse is generally not true, as the function 
    \begin{align}
        g(\epsilon) = f(\epsilon) + e^{-1/\epsilon}
    \end{align}
    is distinct, but has an identical expansion.\footnote{
        This is a consequence of the observation that any derivative of the exponential $e^{-1/\epsilon}$ vanishes at $\epsilon=0$.
    }
    The series expansion can serve as a fingerprint of an analytic function, but is generally not unique. Under a set of mild conditions, Watson's theorem, which is a consequence of analyticity, and its generalisations guarantee that the mapping becomes one-to-one.\footnote{
        Watson's theorem was the first attempt to build a one-to-one correspondence between analytic functions and their expansions \cite{Hardy:1962}. F. Nevanlinna \cite{Nevanlinna:1918} and later Sokal \cite{Sokal:1980} extended Watson's theorem to a more general class of analytic functions. Also, see \cite{Delabaere:2006} for an extension of this work.
        \begin{itemize}
            \item When the function $f$ is analytic in a circle $C_R$ of radius $R$ centred at $R$ in the complex $\epsilon$ plane, and when in addition the residue of the partial sum $R_N(z)$ defined by 
            \begin{align}
                f(\epsilon) = \sum_{n=0}^{N-1} f_n \epsilon^n+ R_N(\epsilon)
            \end{align}
            is bounded by 
            \begin{align}
                |R_N(\epsilon)| \leq A \sigma^N N! |\epsilon|^N\label{eq:residue}
            \end{align}
            for some positive $A$ and $\sigma$, the Borel transform $\phi(t) = \sum_{n=0}^\infty f_n t^n / n!$ converges along a region around the positive real half line, and the function $f$ can be written as the Laplace transform 
            \begin{align}
                f(\epsilon) = \frac{1}{\epsilon} \int_0^\infty e^{-t/\epsilon} \phi(t) \mathrm{d}t\,,\label{eq:Laplace_ex}
            \end{align}
            for $z \in C_R$. In our context $\epsilon = 1/\omega$.
        \end{itemize}
        \begin{itemize}
            \item Conversely, when we can write the function $f$ as a Laplace type integral \eqref{eq:Laplace_ex} with the function $\phi$ analytic in a neighbourhood of the positive real half-line, and moreover, 
            \begin{align}
                |\phi(t)|  \leq K e^{|t|/R}\label{eq:Bound_on_B}
            \end{align}
            for some $K$ in this region, the expansion of the function $f(\epsilon)$ obtained from $f_n = \phi^{(n)}(t)|_{t=0}$ satisfies the property \eqref{eq:residue}. 
        \end{itemize} 
        Under these conditions, the functions $f$ and its expansions $\sum f_n \epsilon^n$ are in a one-to-one correspondence.}
    We conclude that an asymptotic series resulting from the Kirchhoff-Fresnel integral is Borel resumable as we can write it as Laplace-type integral. Since the original integral and the Borel resummed expression share the same expansion and are both of Laplace type, the resummed expression needs to coincide with the original lens integral and the full information of the lens integral beyond the superasymptotic approximation is encoded in the diverging tails of the asymptotic series of the transseries! It turns out that Borel resummation is indeed one method among a large class of resummation methods, constructing a convergent expression with the same asymptotic expansion, and leveraging the correspondence between analytic functions and their asymptotic series (see, for another prominent example, Mittag-Leffler summation \cite{MittagLeffler:1908}). For an insightful review and further details, we refer the reader to \cite{berry2017divergent}.

    \bigskip
    In \cref{sec:Laplace}, we changed coordinates and rewrote the Kirchhoff-Fresnel integral as the Laplace type integral of the algebraic curve $\phi_{\bm{y}}(t)$. The discussion above indicates that $\Delta \phi^{(j)}_{\bm{y}}$ may be interpreted as the Borel transform, \textit{i.e.},
    \begin{align}
        \Delta \phi^{(j)}_{\bm{y}}(t) = \sum_{m=0}^\infty \frac{T_m^{(j)}(\bm{y})}{m!} t^m\,.
    \end{align}
    In practice, one can therefore either Borel transform the asymptotic series or perform the change of coordinates on the integral. Another alternative is to derive the asymptotic series from the algebraic curve associated to the Borel transform by expanding around the branch points (see \cref{ap:algebraic_curve} for further details on this approach).

    In the Laplace representation, the function $\phi_y(t)=(dt/dx)^{-1}$ is multi-valued because the equation $t=-iT(x,y)$ has multiple local inverses. Hence $\phi_y$ defines a Riemann surface over the $t$-plane with branch points at $t_j=-iT(x_j,y)$. For the contribution of a fixed saddle $x_j$, the map $t=-iT$ sends the steepest descent contour $\mathcal{J}_j$ to a branch cut beginning from $t_j$, and the integral along $\mathcal{J}_j$ becomes an integral of the discontinuity across that cut,
    \begin{equation}
    \Delta\phi^{(j)}_y(t):=\phi^{(+)}_{y,j}(t)-\phi^{(-)}_{y,j}(t),
    \end{equation}
    where $\pm$ denote boundary values on the two sides of the cut. Near a saddle point, the local behaviour is
    \begin{equation}
        \Delta\phi^{(j)}_y(t) = \frac{(2\pi i)^{d/2}}{\Gamma(d/2)}\frac{t^{\frac d2-1}}{\sqrt{\det\nabla\xi(x_j)}}\left(1+O(t)\right).
    \end{equation}
    Defining
    \begin{equation}
    U_y^{(j)}(\omega):=\omega^{d/2}\int_0^\infty e^{\omega t}\,\Delta\phi^{(j)}_y(t)\,dt,
    \end{equation}
    and comparing with the Borel expansion
    \begin{equation}
       \Psi^{(j)}(y)\sim\frac{e^{i\omega T(x_j,y)}}{\sqrt{\det\nabla\xi(x_j)}}\sum_{m\ge0}\frac{T_m^{(j)}(y)}{\omega^m},
    \end{equation}
    we see that
    \begin{equation}
    U_m^{(j)}(y) = (2\pi i)^{d/2}\frac{\Gamma(m+\tfrac d2)}{\Gamma(\tfrac d2)}T_m^{(j)}(y),
    \end{equation}
    where $U_m^{(j)}$ denotes the $m\textsuperscript{th}$ term in the $\omega$ expansion of $U_y^{(j)}(\omega)$.

In practical applications one typically has access only to a finite set of coefficients $T_m^{(k)}$ in the Borel transform \eqref{eq:BorelTransform}. Resurgence theory relates the large-order behaviour of these coefficients to
the analytic structure of the Borel transform, in particular to the location and nature of its singularities.
By Darboux's theorem (see, for example, \cite{GauntGutmann1974,Henrici1977}), the late terms of a
power series encode the singularities of its analytic continuation in the Borel plane.

Given finitely many coefficients of a power series, Pad\'e approximation is a practical method to approximate the analytic continuation by a rational function. The poles of the resulting Pad\'e approximant often give accurate numerical estimates of Borel singularities, which correspond to adjacent saddle points in
the Picard-Lefschetz decomposition and ultimately the associated Stokes constants may be determined.
Combined with Borel resummation, this procedure is known as the Pad\'e-Borel method; see for example
\cite{aniceto2019primer,costin2021conformal} for further details and references therein.

While Pad\'e-Borel techniques are widely used in applications across theoretical physics~\cite{marino2022new,aniceto2012resurgence,dondi2020towards,lustri2023exponential}, their rigorous convergence properties remain subtle and problem-dependent. In the present work we instead make use of hyperasymptotic methods, which allow us
to extract Stokes constants and provides controllable error bounds that are particularly explicit in the case of integral problems.

    \subsection{The hyperasymptotic approximation}\label{sec:hyperasymptotics}
    Borel resummation of the asymptotic series is a theoretical construction that demonstrates that the lens integral can be reconstructed from the transseries. However, it does not offer a practical means of converting the divergent tails into systematic corrections beyond the superasymptotic level for the majority of problems in wave optics and physics in general. For example, for Borel resummation to be an effective method, one must first know all terms of the asymptotic series and subsequently compute its Borel transform by recognising it as some known special function. If the Borel transform is truncated, the subsequent Laplace transform merely reproduces the original asymptotic series. In contrast, hyperasymptotics is a systematic framework to extract finite approximations (with error bounds) from a formal divergent transseries. The hyperasymptotic approximations require only finitely many terms from the full transseries. Below, we review hyperasymptotics, highlighting the key points following the derivation by Berry and Howls \cite{Berry:1991, Howls:1997}.

\begin{figure}
        \centering
        \includegraphics[width=0.5\textwidth]{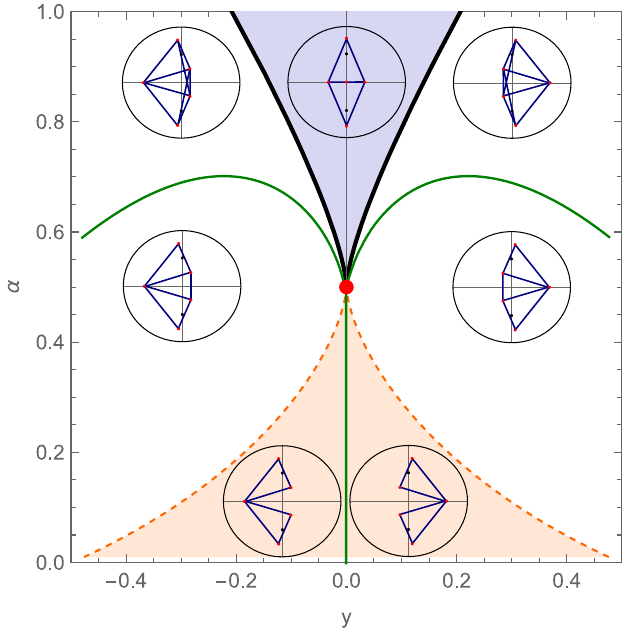}
        \caption{Caustics (black) with the cusp (red), Stokes lines (blue) and higher-order Stokes lines (green) of the Lorentzian lens in the $y$-$\alpha$ plane. The corresponding adjacency graphs are shown in the circled diagrams displaying the complex $x$-plane. The adjacency graphs change at the caustics, Stokes lines and higher-order Stokes lines.}\label{fig:caustics_Higherorder}
    \end{figure}

    Starting with the Picard-Lefschetz formula in the Laplace representation, we study the integral along the steepest descent manifold $\mathcal{J}_j$ associated with the classical ray $\bm{x}_j$ (\cref{eq:Laplace_nD}),
    \begin{align}
        \Psi^{(j)}(\bm{y}) 
        &=\left(\frac{\omega}{2 \pi i}\right)^{d/2} \int_{\mathcal{J}_j} e^{i \omega T(\bm{x},\bm{y})}\mathrm{d}\bm{x}\\
        &= \frac{e^{i \omega T(\bm{x}_j,\bm{y})}}{\left(2\pi i\right)^{d/2}}U^{(j)}_{\bm{y}}(\omega)\,,
    \end{align}
    with 
    \begin{align}
        U^{(j)}_{\bm{y}}(\omega) = \omega^{d/2} \int_{0}^\infty e^{-\omega t}\Delta \phi_{\bm{y}}^{(j)}(t)\mathrm{d}t\,,
    \end{align}
    and its asymptotic series of the form $U^{(j)}_{\bm{y}}(\omega) = \sum_{m=0}^\infty \frac{U^{(j)}_m}{\omega^m}$ where $U^{(j)}_m=\frac{(2\pi i)^{d/2}}{ \sqrt{\det \nabla \bm{\xi}(\bm{x}_j)}}  T_m^{(j)}(\bm{y})$. The function $U^{(j)}_{\bm{y}}$ and the coefficients $U_m^{(j)}$ correspond to $T^{(j)}_{\bm{y}}$ and $T_m^{(j)}$ in the notation of Berry and Howls \cite{Berry:1991,Howls:1997}.
    Using Cauchy's residue theorem, we can write 
    \begin{align}
        t^{1-d/2} \Delta\phi_{\bm{y}}^{(j)}(t) = \frac{1}{2 \pi i} \oint_{\bar{\gamma}(t)} \frac{\Delta\phi_{\bm{y}}^{(j)}(\zeta) \zeta^{1-d/2}}{\zeta -t}\mathrm{d}\zeta    \,,
    \end{align}
    with $\bar{\gamma}(t)$ a loop around $t$ in the Borel plane. Consequently, the Laplace-type integral can be written as
    \begin{align}
        U^{(j)}_{\bm{y}}(\omega) = \frac{\omega^{d/2}}{2 \pi i} \int_{0}^\infty e^{-\omega t} t^{d/2-1}\left[  \oint_{\bar{\gamma}(t)} \frac{\Delta\phi_{\bm{y}}^{(j)}(\zeta) \zeta^{1-d/2}}{\zeta -t}\mathrm{d}\zeta\right]\mathrm{d}t\,,
    \end{align}
    where $\Gamma_j$ (known as the sausage contour in the Borel plane in \cite{Berry:1991}) is a loop around the steepest descent manifold $\mathcal{B}\mathcal{J}_j$ in the Borel plane (the union of $\bar{\gamma}(t)$ for all $t$ on $\mathcal{B}\mathcal{J}_j$). So far, it appears that we have only complicated the integral. However, substituting the expansion 
    \begin{align}
        \frac{1}{1-x} = \sum_{r=0}^{N-1} x^r + \frac{x^N}{1-x}\,,\label{eq:expansion}
    \end{align}
    we recover the truncated asymptotic series and its associated error term
    \begin{align}
        U^{(j)}_{\bm{y}}(\omega) = \sum_{m=0}^{N-1} \frac{U^{(j)}_m}{\omega^m} + \frac{\omega^{d/2}}{2 \pi i} \int_0^\infty e^{-\omega t} t^{N+d/2-1} \left[\oint_{\Gamma_n} \frac{\Delta \phi^{(j)}(\zeta)}{\zeta^{N+d/2}(1-t/\zeta)}\mathrm{d}\zeta\right]\mathrm{d}t
    \end{align}
    A detailed analysis shows that the optimal truncation $N$, minimising the error term, is given by the integer part of the distance to the closest saddle point in the Borel plane, \textit{i.e.}, the integer part of $|\omega F_{jk^*}|$, giving rise to the superasymptotic approximation. The error is of the order $e^{-|\omega F_{jk^*}|}$. The so-called resurgence relation is obtained by realising that the error term is a Laplace-type integral and that the contour $\Gamma_j$ can be deformed onto a set of steepest descent contours in the Borel plane that live on the same Riemann sheet (and can be reached with straight lines from the branch point $t_j$ in the Borel plane). A saddle point, whose descent manifold participates in this deformation, is known as adjacent. Expanding the integrals along the adjacent thimbles in the Borel plane, we obtain the resurgence relation
    \begin{align}
        U^{(j)}_{\bm{y}}(\omega) = \sum_{m=0}^{N-1}\frac{U^{(j)}_m}{\omega^m} + \frac{1}{2 \pi i} \sum_k \frac{K_{jk}}{(\omega F_{jk})^N}\int_0^\infty \frac{e^{-\nu} \nu^{N-1}}{(1-\nu / \omega F_{jk})} U^{(k)}_{\bm{y}}\left(\frac{\nu}{F_{jk}}\right)\,,
    \end{align}
    with the singulant
    \begin{align}
        F_{jk} = i(T(\bm{x}_j,\bm{y})-T(\bm{x}_k,\bm{y}))\,.
    \end{align}
    The sum ranges of the adjacent rays to $\bm{x}_j$ for which the so-called Stokes constants $K_{jk} \in \{-1,0,1\}$ do not vanish. The Stokes constants $K_{ij}$ capture the Riemann sheet structure of the Borel transform and are neatly represented by the adjacency graph, where the branch points $t_j$ and $t_k$ in the Borel plane are connected by a line when $|K_{ij}| \neq 0$ (see \cref{fig:caustics_Higherorder} for the adjacency graph for the one-dimensional Lorentzian lens as a function of the lens amplitude $\alpha$ and the point on the image plane $\bm{y}$). Two saddle points may be non-adjacent when they live on different Riemann sheets of the Borel transform. Their ``sight" is obstructed by a branch point. For one-dimensional integrals, the adjacency may be inferred by varying the phase of the frequency $\omega$. When two rays undergo a Stokes phenomenon as we change the phase, the rays are adjacent (see the discussion by Berry and Howls \cite{Berry:1991} and the ``radar method" by Voros \cite{Voros:1983}). For multidimensional integrals, Olde Daalhuis developed a method to compute the adjacency relations from the asymptotic series \cite{Daalhuis:1999}. The adjacency graph changes as we vary $\alpha$ and $\bm{y}$ while the saddle points and associated branch points move in the complex $\bm{x}$- and $t$-planes and change relevance (see \cref{fig:caustics_Higherorder}). Caustics and Stokes' phenomena mark a change in the relevance of a classical ray. Analogously, when the adjacency graph changes, the geometry of the Riemann sheet undergoes a qualitative change known in a higher-order Stokes phenomenon \cite{Howls:2004,Howls:2025,shelton2025exponential}. 

    The resurgence relation implies the following relation between the coefficients of the different asymptotic series
    \begin{align}
        U_N^{(j)} \approx \frac{1}{2 \pi i} \sum_{m} \sum_{r=0}^{N_m} \frac{(N-r-1)!}{F_{nm}^{N-s}}U_s^{(m)}\,,
    \end{align}
    where the sum is taken over adjacent saddles. The leading contribution comes from the closest adjacent ray in the Borel plane,
    \begin{align}
        U_r^{(j)} \approx U_0^{(k^*)} \frac{K_{jk^*}(r-1)!}{F_{jk^*}^r}\,,
    \end{align}
    with the label $k^*$ marking the adjacent saddle point $\bm{x}_{k^*}$ with the smallest singulant $|F_{jk^*}|$.
    The fact that we can reconstruct the presence, the exponential and the asymptotic series from an adjacent ray is the origin of the term ``resurgence". In \cref{sec:physicalImpliations}, we indicate how the resurgence relation leads to non-trivial relations between lensed waveforms and can, in principle, be used to reconstruct the phase variation from observations beyond the methods based on the geometric optics or eikonal approximation. Resurgence is ultimately an application of analyticity, built on the fact that the Taylor series of an analytic function converges on a finite disk and repeated expansions define the analytic continuation. We illustrate the resurgence relation with an example.

\begin{figure}
        \centering
        \includegraphics[width=0.5\textwidth]{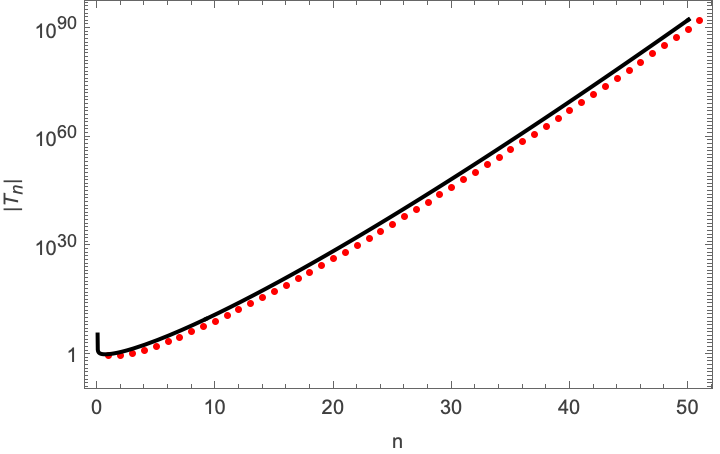}
        \caption{The scaling of the coefficients of the saddle point $x=0$ of the quartic integral (the red points) reveals the presence of saddle points at $x=\pm i/\sqrt{2}$ as can be seen from the correspondence with the function $f(n)=\frac{1}{\sqrt{2}}\frac{\Gamma(n)}{(1/4)^n}$ (the black curve)}\label{fig:resurgence}
    \end{figure}

    \begin{example}
        In \cref{ex:quartic_trans}, we derived the transseries 
        \begin{align}
            \Phi \sim n_{x=0} \sum_{n=0}^\infty \frac{(4 n - 1)!!}{4^{ n} n! i^n \omega^n}  + \left(n_{x=- i / \sqrt{2}} + n_{x= i / \sqrt{2}}\right) \frac{e^{-\frac{i \omega}{4}} }{\sqrt{2} i } \left[1-\frac{3i}{16 \omega^2} + \frac{15 i}{8 \omega^3} -\frac{105}{256 \omega^4} + \dots \right]\,,
        \end{align}
        for the quartic integral 
        \begin{align}
            \Phi = \sqrt{\frac{\omega}{\pi i}}\int_{-\infty}^\infty e^{i \omega (x^2 + x^4)}\mathrm{d}x\,.
        \end{align}
        Plotting the coefficients $T_n = \frac{(4 n - 1)!!}{4^{ n} n! i^n}$ we observe the presence of the other saddle points with the exponential factor $-i/4$, through the scaling $\frac{4^n \Gamma(n)}{\sqrt{2}} $ (see \cref{fig:resurgence}).
    \end{example}

    We consider applying the resurgence relation \eqref{eq:expansion} to the integral representation for the error term and iterating this procedure. In this way, one obtains the so-called hyperasymptotic expansion 
    \begin{align}
        U^{(n)}_{\bm{y}}(k) &= 
        \sum_{r=0}^{N_n - 1} U_r^{(n)} K_r^{(n)}(k) + 
        \sum_{m_1=1} K_{n m_1} \sum_{r=0}^{N_{n m_1}-1} U_r^{(m_1)}K_r^{(nm_1)}(k) \nonumber \\
        &\phantom{=}+ \sum_{m_1=1}\sum_{m_2=1} K_{n m_1} K_{m_1 m_2}\sum_{r=0}^{N_{n m_1 m_2}-1} U_r^{(m_2)}K_r^{(nm_1m_2)}(k) \nonumber\\
        &\phantom{=}+ \dots \nonumber\\
        &\phantom{=}+ \sum_{m_1=1} \cdots \sum_{m_M=1} K_{n m_1} \cdots K_{m_{M-1} m_M}\sum_{r=0}^{N_{n m_1 \cdots m_M}-1} U_r^{(m_M)}K_r^{(nm_1 \cdots m_M)}(k)\nonumber\\
        &\phantom{=}+ \sum_{m_1=1} \cdots \sum_{m_M=1} K_{n m_1} \cdots K_{m_{M-1} m_M} R^{(n m_1 \cdots m_M)}\,,
    \end{align}
    where the terms
    \begin{align}
        K_r^{(n)}(k) & = \frac{1}{k^r}\,,\\
        K_r^{(n m_1)}(k) 
        & = \frac{1}{2 \pi i k^{N_{n}-1}} \int_0^{e^{i \theta}\infty} \frac{t^{N_{n}-r-1} e^{-F_{n m_1} t}}{k - t}  \mathrm{d}t \nonumber\\
        & = \frac{(-1)^{N_n -r} e^{-k F_{nm}}  \Gamma(N_n -r) \Gamma(1- N_n+r, - k F_{nm})}{2 \pi i k^{r}}\,,\\
        K_r^{(n m_1 \dots m_M)}(k) & = 
        \frac{1}{(2 \pi i )^{M+1} k^{N_{n}-1} k^{N_{m_1}-1} \cdots k^{N_{m_M}-1}} \int_0^{e^{i \theta_0}\infty} \cdots \int_0^{e^{i \theta_M}\infty} \mathrm{d}t_M \cdots \mathrm{d}t_0\nonumber\\
        \phantom{=}& \times  \frac{t_0^{N_n - N_{m_1}-1}\cdots t_{M-1}^{N_{m_{M-1}}-N_{m_M}-1}t_M^{N_{m_M}-r-1} e^{-F_{n m_1} t_0- F_{m_1 m_2} t_1-\cdots - F_{m_{M-1} m_M} t_M}}{(k - t_0)(t_0 - t_1)\cdots (t_{M-1} - t_M)}  \,.
    \end{align}
    are known as hyperterminants and were first introduced in \cite{Berry:1990,Berry:1991}. Soon afterwards, Olde Daalhuis proposed an efficient way to evaluate them using convergent series \cite{Daalhuis:1996,Daalhuis:1998}. The optimal truncation points
    \begin{align}
        N_n &= \{\text{shortest path for $M$ steps in the Borel plane starting from $\bm{x}_n$}\}\,,\\
        N_{nm_1} &= \max\{0, N_n - |\omega F_{nm_1}|\}\,,\\
        &\vdots \nonumber\\
        N_{nm_1m_2 \cdots m_M} &= \max\{0, N_{nm_1 \cdots m_M} - |\omega F_{m_{M-1} m_M}|\}\,,
    \end{align}
   minimising the error
    \begin{align}
        |R^{(nm_1 \cdots m_M)}| \approx& \frac{N_{n\cdots m_M}^{N_{n\cdots m_M}-1/2} \exp(-N_n)}{(2\pi)^{(M+1)/2}|k|^{N_n} |F_{m_M m_{M+1}^*}|^{N_{n \cdots m_M}}}\nonumber\\
        &\times \prod_{p=0}^{M-1} \frac{(N_{n \cdots m_p} - N_{n\cdots m_p m_{p+1}})^{(N_{n \cdots m_p} - N_{n\cdots m_p m_{p+1}}-1/2)}}{|F_{m_p m_{p+1}}|^{N_{n \cdots m_p} - N_{n \cdots m_p m_{p+1}}}} |U_0^{(m_{M+1}^*)}|\,.
    \end{align}
    were derived in \cite{Daalhuis:1998c} and reproduced in \cite{Howls:1997}. The hyperasymptotic approximation of the steepest descent integral $\Psi^{(j)}(\bm{y})$, starts by including the first $N_n-1$ of the asymptotic series. The next level of the approximation is based on the first $N_{nm_1}-1$ terms of this asymptotic series of the adjacent ray $m_1$, scaled by the first-order hyperterminants. At every subsequent level, more adjacent rays contribute. Graphically, the hyperasymptotic approximation follows paths in the adjacency graph, increasing the accuracy of the approximation at every level. By moving beyond the truncation point of the superasymptotic approximation, into part of the diverging tail of the expansion, the hyperasymptotic approximation achieves an exponential improvement at every level. Consequently, the refractive expansion performs best in the refractive regime but may, in principle, be used to explore the diffractive regime as well.

    We illustrate the hyperasymptotic approximation for the one-dimensional Lorentzian lens in \cref{fig:hyper_omega}. The first-order hyperasymptotic approximation is a significant advance over the superasymptotic approximation away from the caustics and manages to recover the lens integral with greater accuracy near the caustics. In the direct vicinity of the caustics, the hyperasymptotic expansion of the nondegenerate transseries reduces to the eikonal approximation. The hyperasymptotic approximation also enables us to model the lens integral for smaller frequencies than the superasymptotic approximation. By increasing the level of the hyperasymptotic approximation, we can probe the lens integral with great levels of accuracy approaching the diffractive regime. In an upcoming paper, we will publish a numerical library to efficiently perform hyperasymptotic approximations of oscillatory integrals.

    \subsubsection{Physical implications}\label{sec:physicalImpliations}
    Finally, the resurgence of rays in the asymptotic series demonstrates how adjacent classical rays -- not necessarily relevant in the Picard-Lefschetz theory sense -- influence the lens amplitude. Formally, if we were to measure the lens amplitude $\Psi(\bm{y})$ to a great level of accuracy as a function of the frequency $\omega$ for fixed $\bm{y}$, we could, in principle, reconstruct the adjacent rays and reconstruct the phase variation. This should be contrasted with the eikonal approximation, which relates the lens amplitude solely to the deformation tensor $\nabla \bm{\xi}$ of the relevant images (corresponding to the Hessian of the phase variation $\varphi$) and treats the different images as being independent.

\begin{figure}
        \centering
        \begin{subfigure}[b]{0.45\textwidth}
            \includegraphics[width=\textwidth]{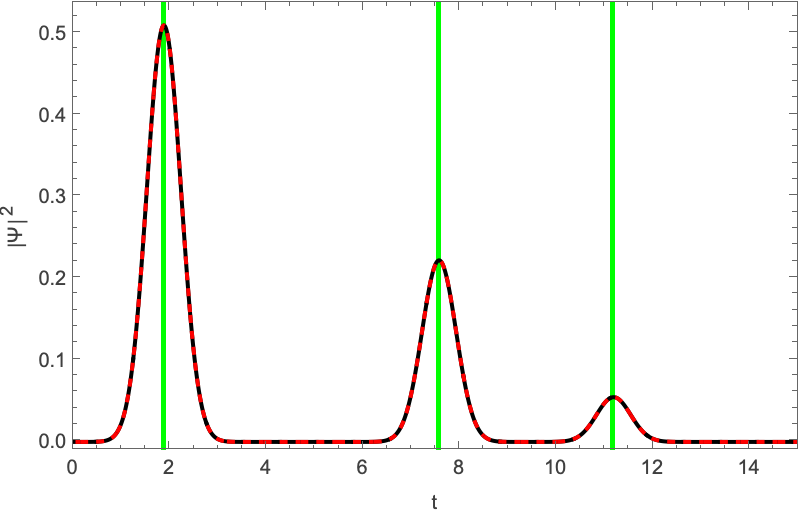}
            \caption{The waveform.}
        \end{subfigure}
        \begin{subfigure}[b]{0.45\textwidth}
            \includegraphics[width=\textwidth]{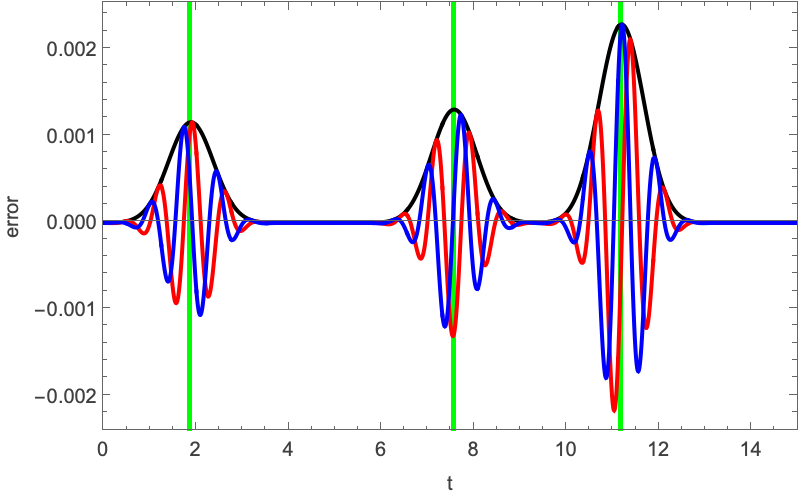}
            \caption{The error}
        \end{subfigure}
        \caption{\textit{Left:} The lensed waveform $|F(t)|^2$ of a Gaussian lens with $\mathcal{A}=1, \sigma =\frac{1}{2},\mu =10$ by the one-dimensional Lorentzian lens with the lens strength $\alpha = 10$ and position $y = 3/2$ (the black curve) and the associated eikonal approximation $F_{\text{eikonal}}(t)$ (the red curves) as a function of time. \textit{Right:} the difference between the numerical and eikonal evaluation of the waveform, with the real, imaginary part and modulus of $F(t)$  and $F_{\text{eikonal}}(t)$.}\label{fig:waveform}
    \end{figure}

    As a thought experiment, let us consider a coherent source of long-wavelength radiation that emits a Gaussian waveform 
    \begin{align}
        f(t) = \mathcal{A}\, e^{-\frac{t^2}{2 \sigma ^2}+i \mu  t}
    \end{align}
    with the mean frequency of the radiation $\mu$, the spread of the pulse $\sigma$ and the amplitude $\mathcal{A}$, and analyse the resurgent information in the waveform. Can one, in principle, predict the occurrence of an image from an observation of a lensed image? Using the convolution theorem, we model the lensed waveform as a Fourier integral over the lens amplitude
    \begin{align}
        F(t) = \int \Psi(\bm{y}) \hat{f}(\omega) e^{-i \omega t}\mathrm{d}\omega\,,
    \end{align}
    with $\hat{f}(\omega) = \frac{\sigma\, \mathcal{A} }{\sqrt{2 \pi }} e^{-\frac{1}{2} \sigma ^2 (\mu +\omega )^2}$ the Fourier transform of the original waveform $f$. When $\bm{y}$ resides in a multi-image region, the waveform $F(t)$ consists of several lensed copies of the original waveform displaced by the time delay of the real classical rays (see, for example, \cref{fig:waveform}). In the eikonal approximation, including both the real and complex relevant classical rays, the lensed waveform is given by
    \begin{align}
        F_{\text{eikonal}}(t) 
        &=\sum_j  \frac{n_j(\bm{y})}{\sqrt{\det \nabla \bm{\xi}(\bm{x}_j)}} 
        \int  \hat{f}(\omega)  e^{-i \omega (t - T(\bm{x}_j,\bm{y})) } \mathrm{d}\omega\\
        &=\sum_j  \frac{n_j(\bm{y})}{\sqrt{\det \nabla \bm{\xi}(\bm{x}_j)}} 
        f(t-T(\bm{x}_j,\bm{y}))\,.
    \end{align}
    Given the detailed observation of one of the waveforms, we could, in principle, predict the images that are still to arrive at the telescope. For the lensed waveform in \cref{fig:waveform}, it is clear that the eikonal approximation is very accurate, and that the deviations that formally capture the resurgent information about the adjacent classical rays are small. The deviations become larger for smaller frequencies and near caustics, revealing more of the resurgent information in the waveform. However, in both instances, the different lensed images start to overlap and remain unresolved. In fact, this is an example of Bohr’s principle of complementarity. Like Young’s experiment, one cannot both see the interference fringes and tell which slit the photon passed through. Remarkably, in the lensing of coherent astrophysical point sources, like pulsars, fast radio bursts, and gravitational waves, such effects operate over galactic, or even cosmological, scales.  In this case, it becomes increasingly difficult to predict a future image with resurgence when the present image is fully resolved.

    \section{Conclusion}\label{sec:conclusion}  

    Wave optics has become a new coherent window to the universe, probed by gravitational waves, Fast Radio Bursts and more.  New physical observables arise, which have traditionally only been studied in the refractive (eikonal) or diffractive (perturbative) limits.  This work generalizes the concepts, opening potential for new measurements. This paper uses the Lorenzian lens as a template to quantify wave properties of caustics, which have traditionally escaped analyses.

    In this paper, we explore analytic approximations of lensing in wave optics, extending earlier work by \cite{Jow:2023}. We introduce resurgence theory to astrophysical lensing, develop these methods for multi-dimensional oscillatory integrals, including new algorithms to evaluate the asymptotic series. We aim to provide a pedagogical introduction to both transseries, Borel resummation, superasymptotic and hyperasympotic approximations. This work sheds new light on the lensing of coherent sources in wave optics. This is particularly timely given the recent observations of pulsars, fast radio bursts and gravitational waves. The techniques presented in this paper apply to oscillatory integrals in general. In the future, we aim to apply them to the study of real-time Feynman path integrals.

    To our surprise, we demonstrate that -- despite Dyson's argument for the emergence of asymptotic series -- the diffractive expansion of the Kichhoff-Fresnel integral converges for the vast majority of physical lenses. The convergence is irrespective of the frequency and formally works in both the diffractive and the refractive regimes. Though in the refractive regime, the convergence emerges from the cancellation of large terms and can become computationally expensive. Moreover, we provide a new method to study lensing in wave optics by means of the diffractive expansion of the Gaussian lens. Remarkably, this method does not require any knowledge of classical rays or caustics and scales well with the dimensionality of the lens integral. As this method scales well with the dimension of the oscillatory integral, we expect that the convergence of the diffractive expansion and its implementation in terms of Gaussian lenses can aid the exploration of problems beyond wave optics.

    Next, we introduce the study of the refractive expansion, going beyond the eikonal expansion, and introduce the mathematical framework of resurgence. We show that the refractive expansion gives rise to a transseries consisting of divergent asymptotic expansions. Moreover, we provide a new algorithm for the evaluation of the asymptotic expansions associated with multidimensional integrals. Near caustics, we provide a practical method to evaluate the uniform asymptotic expansion near caustics, going beyond earlier studies in wave optics \cite{Grillo:2018}. Finally, we demonstrate how these transseries can be transformed into ever more accurate approximations of the lens amplitude. Our study illustrates that the refractive expansion is most efficient in the refractive regime, but can, in principle, be used to explore the diffractive regime as well. The refractive expansion and the resurgence analysis provide a deep understanding of the interrelations of classical rays through complex analysis.

\acknowledgments

  SC thanks Heng-Yu Chen for the invitation to visit National Taiwan University. During this visit SC met ULP and was introduced to JF, which led to the present work. We thank Dylan Jow for collaboration at an early stage of the project. We are particularly grateful also to In\^es Aniceto, Gerg\H{o} Nemes, Adri Olde Daalhuis, and Chris Howls for many insightful discussions on asymptotics and resurgence.The work of SC is supported by National Science and Technology Council of Taiwan under Grants No. NSTC 113-2112-M-007-019, NSTC 114-2112-M-007-015

    The work of JF is supported by the STFC Consolidated Grant ‘Particle Physics at the Higgs Centre,’ and, respectively, by a Higgs Fellowship at the University of Edinburgh.


\bibliographystyle{JHEP}
\bibliography{library.bib}
    \appendix

    \section{Algebraic curves}\label{ap:algebraic_curve}
    In \cref{sec:Laplace}, we obtain the Borel transform by means of a change of coordinates, $t=-i T(x,y)$, through the Jacobian factor
    \begin{align}
        \phi_y(t) = \left[\frac{\mathrm{d}t}{\mathrm{d}x}\right]^{-1}\,.
    \end{align} 
    To derive $\phi_y(t)$ in practice, we first evaluate the derivative $\left[\frac{\mathrm{d}t}{\mathrm{d}x}\right]^{-1} = \frac{i}{x-y + \varphi'(x)}$, next solve $t=-i T(x,y)$ for $x$ and finally substitute $x(t)$ to obtain $\phi_y$. However, as $t=-iT(x,y)$ often has multiple solutions for $x(t)$, the Borel transform $\phi_y$ is multivalued with branch point singularities located at the time delay of the classical rays. Moreover, it is not always possible to find $x(t)$ analytically. We will here restrict our analysis to rational lens models and interpret the Borel transform $\phi_y$ as a Riemann surface on the Borel plane defined as an algebraic curve. The algebraic curve prescription allows us to probe $\phi_y$ implicitly when the explicit solution $x(t)$ cannot be found or is too complicated to work with.

\begin{figure}
        \centering
        \begin{subfigure}[b]{0.45\textwidth}
            \includegraphics[width=\textwidth]{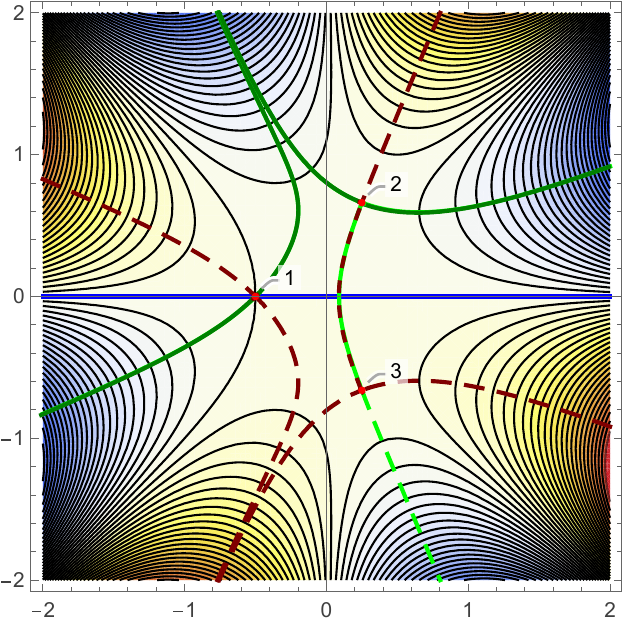}
            \caption{The complex $x$-plane.}
        \end{subfigure}
            \begin{subfigure}[b]{0.45\textwidth}
        \includegraphics[width=\textwidth]{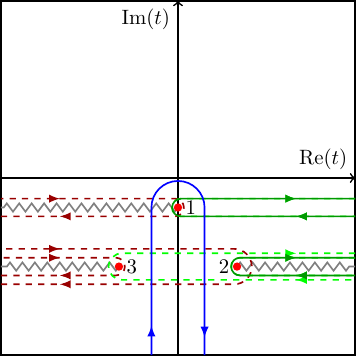}
        \caption{The Borel plane.}
    \end{subfigure}
        \caption{The Picard-Lefschetz deformation of the integral for the quartic lens $\Psi(y) = \sqrt{\frac{\omega}{2 \pi i}} \int \exp\left[i \omega\left(\frac{(x-y)^2}{2} + x^4\right)\right]\mathrm{d}x$ for $y=-1$. \textit{Left:} The complex deformation in the complex $x$-plane, with the original integration domain (blue), the steepest descent manifolds (dashed green), the steepest ascent manifolds (dashed red) and the Picard-Lefschetz deformation (dark green). \textit{Right:} The associated contours in the Borel plane on the Riemann sheets of the Borel transform $\phi_y$, with the branch cuts indicated by the zigzag lines.}\label{fig:quarticLens}
    \end{figure}

    For convenience, we will limit our discussion to the quartic lens with $\varphi(x) = x^4$ with the Kirchhoff-Fresnel integral 
    \begin{align}
        \Psi(y) = \sqrt{\frac{\omega}{2 \pi i}} \int e^{i \omega\left(\frac{(x-y)^2}{2} + x^4\right)}\mathrm{d}x\,.
    \end{align}
    However, the method works for any rational lens model. First note that 
    \begin{align}
        \phi_y(t) = \left[\frac{\mathrm{d} t}{\mathrm{d}x }\right]^{-1} = \frac{i}{x - y + 4 x^3}
    \end{align}
    To construct the algebraic curve, consider the polynomial ring 
    \begin{align}
        \mathbb{C}[x] / (t + i T(x,y))
    \end{align}
    consisting of complex-valued polynomials, where two polynomials are considered equivalent when they differ by a factor $t+i T(x,y)$. Or equivalently, we can identify the highest power in the time delay $x^4$ with the factor $i t- \frac{(x-y)^2}{2}$. Let us assume that $\phi_y$ is the solution of the polynomial identity
    \begin{align}
        0 &= \phi_y^{-4} + a_1 \phi_y^{-3} + a_2 \phi_y^{-2} + a_3 \phi_y^{-1} + a_4\\
        &= (x - y + 4 x^3)^4 + i a_1 (x - y + 4 x^3)^3 - a_2 (x - y + 4 x^3)^2 -i a_3 (x - y + 4 x^3) + a_4\,,
    \end{align}
    for a yet to be determined set of parameters $a_1,a_2,a_3$ and $a_4$. Expanding the brackets and replacing the quartic $x^{4}$ by $it - \frac{(x-y)^2}{2}$, reduce the polynomial till it is of cubic order, \textit{i.e.}, it assumes the form $0 = A_0 + A_1 x + A_2 x^2 + A_3 x^3$. Solving $A_i=0$ for $a_i$ yields the algebraic curve,
    \begin{align}
        0 &=1 + a_1 \phi_y + a_2 \phi_y^2+ a_3 \phi_y^3 + a_4 \phi_y^4\\
        &=1 + 8 i y \phi_y + \left(-8 i t-14 y^2-\frac{1}{2}\right) \phi_y^2 \nonumber\\
        &+ \left((1-192 i t) y^4-8 t (48 t+5 i) y^2+i t (-16 t+i)^2+32 y^6\right) \phi_y^4\,.
    \end{align}
    This curve is considered a locus in the $(\phi_y,t)$ plane with parametric dependence on $y$. The curve is visualised in \cref{fig:quarticLens}. In the complex $x$-plane, we find three saddle points that are mapped to singular branch points in the Borel plane. The original integration domain is mapped to a contour following the imaginary axes from $-i \infty$ to the branch point $1$ after which it turns around and returns to $-i \infty$ on a different Riemann sheet. The steepest ascent and descent contours are mapped to horizontal lines in the complex plane. By intersecting the original integration domain with the steepest ascent contours, we find that saddle point $2$ is relevant and saddle point $3$ is irrelevant to the quartic lens integral. 

    It is possible to infer both the branch points and the asymptotic series of the transseries directly from the algebraic curve prescription. For a more detailed exposition of this method, we refer to \cite{aniceto2024algebraic}.

\end{document}